\documentclass[aps,pra,reprint,groupedaddress,superscriptaddress,floatfix,longbibliography]{revtex4-2}

\usepackage[T1]{fontenc}
\usepackage{newtxtext}
\usepackage{newtxmath}

\usepackage{amsmath}
\usepackage{mathtools}
\usepackage{amsfonts}
\usepackage{bbold}
\usepackage{braket}

\usepackage{color}
\usepackage{xcolor}
\usepackage[colorlinks,citecolor=blue,linkcolor=blue,urlcolor=blue]{hyperref}

\usepackage{graphicx}
\usepackage[all]{hypcap}

\begin{document}

\title{Self-consistent mean-field quantum approximate optimization}

\author{Maxime Dupont}
\email[]{mdupont@rigetti.com}
\affiliation{Rigetti Computing, 775 Heinz Avenue, Berkeley, California 94710, USA}

\author{Bhuvanesh Sundar}
\affiliation{Rigetti Computing, 775 Heinz Avenue, Berkeley, California 94710, USA}

\author{Meenambika Gowrishankar}
\affiliation{Rigetti Computing, 775 Heinz Avenue, Berkeley, California 94710, USA}

\begin{abstract}
    We introduce a self-consistent mean-field quantum optimization algorithm that approximates the ground state of classical Ising Hamiltonians. The algorithm decomposes the problem into independent subproblems and treats the interactions between them in a mean-field manner. These interactions are captured by a common environment---constructed self-consistently through a variational quantum circuit---which modifies the subproblems to account for mutual influence while maintaining computational independence. Consequently, subproblems can be solved individually, avoiding the computational cost of the full problem. We explore the properties of the generated environment and assess the algorithm's performance through extensive numerical simulations on Sherrington-Kirkpatrick spin glasses. Furthermore, we apply it experimentally to a weighted maximum clique problem applied to molecular docking. This framework enables the solution of problems that would otherwise exceed the qubit and gate counts of current quantum hardware.
\end{abstract}

\maketitle

\section{Introduction}

Quantum computing harnesses the laws of quantum mechanics to establish a fundamentally new computational paradigm~\cite{AndrewSteane_1998,Ladd2010}. For specific problem classes, it promises to outperform the most efficient classical algorithms, offering advantages that range from polynomial to exponential speedups. Notable examples of this potential include Shor's algorithm for integer factorization~\cite{shor1997polynomial} and quantum phase estimation for simulating physical systems~\cite{kitaev1995quantum,cleve1998quantum}.

Alternatively, other quantum algorithms are designed as heuristics, relying on empirical results to establish their performance. Variational circuits are the most prominent examples, finding extensive use in quantum machine learning~\cite{cerezo2022challenges} and optimization~\cite{Farhi2014,BLEKOS20241}. However, demonstrating their effectiveness on large-scale problems where classical methods fail remains a significant hurdle. The limitations of contemporary hardware---specifically restricted qubit counts and inherent system noise---severely constrain the circuit sizes that can be reliably executed. This bottleneck prevents the execution of algorithms at the scale necessary to potentially achieve a practical quantum advantage in real-world scenarios.

In the domain of optimization, claims of asymptotic scaling advantage have been made, yet their practical demonstration necessitates significant improvements in device scale and fidelity. Recent studies have provided compelling evidence: Ref.~\cite{Ebadi2022} reported an experimental scaling advantage over simulated annealing for maximum independent set problems on a $289$-atom platform. Similarly, numerical extrapolations suggest that the quantum approximate optimization algorithm (QAOA)~\cite{Farhi2014} at depth $p\geq 11$ offers an advantage for low autocorrelation binary sequences~\cite{shaydulin2024evidence}, while Ref.~\cite{Cadavid2025} found that quantum enhancements to state-of-the-art classical solvers yield a scaling benefit on the same problem. Analogous scaling analyses have been conducted using D-Wave's quantum annealer against various classical benchmarks~\cite{PhysRevX.8.031016,Bauza2024}. While establishing an absolute runtime advantage remains challenging amidst rapidly evolving classical benchmarks~\cite{Chandarana2025,Chandarana2025b,Farre2025,Tuziemski2025}, the collective evidence hints that as quantum processors scale in size and fidelity, they may unlock a genuine practical advantage in optimization.

Although hardware advancement remains pivotal for scaling circuit executions, algorithmic innovations have emerged to address optimization instances that exceed the native capacity of current devices. Strategies to mitigate resource constraints include novel encoding schemes~\cite{patti2022variational,Tenecohen2023,PhysRevResearch.5.L012021,PhysRevA.109.052441,PhysRevResearch.4.033142,Sciorilli2024,Sundar2024}, circuit-cutting techniques~\cite{Bechtold_2023}, and methods that exploit the causal light-cone structure of the QAOA~\cite{PhysRevApplied.23.014045,9prw-684p}. Moreover, decomposition approaches have gained prominence; these techniques reduce complexity by partitioning the global problem into manageable subproblems~\cite{Liu2022,tomesh2023divide,zhou2023qaoa,vcepaite2025quantum,Huang2026}. Other recent approaches have focused on graph-theoretic techniques, employing decomposition and sparsification to reduce problem complexity and mitigate noise in variational algorithms~\cite{Ponce2023,Moondra2024,Acharya2024}. A distinct approach is found in multilevel decomposition schemes, which tackle complexity through hierarchical coarsening rather than simple partitioning. By iteratively reducing the problem size and refining solutions across levels, these frameworks allow solvers to successfully address large-scale problems that far exceed native device capacities~\cite{10363584,Bach2024,Maciejewski2024}.

\begin{figure}[!t]
    \centering
    \includegraphics[width=1\columnwidth]{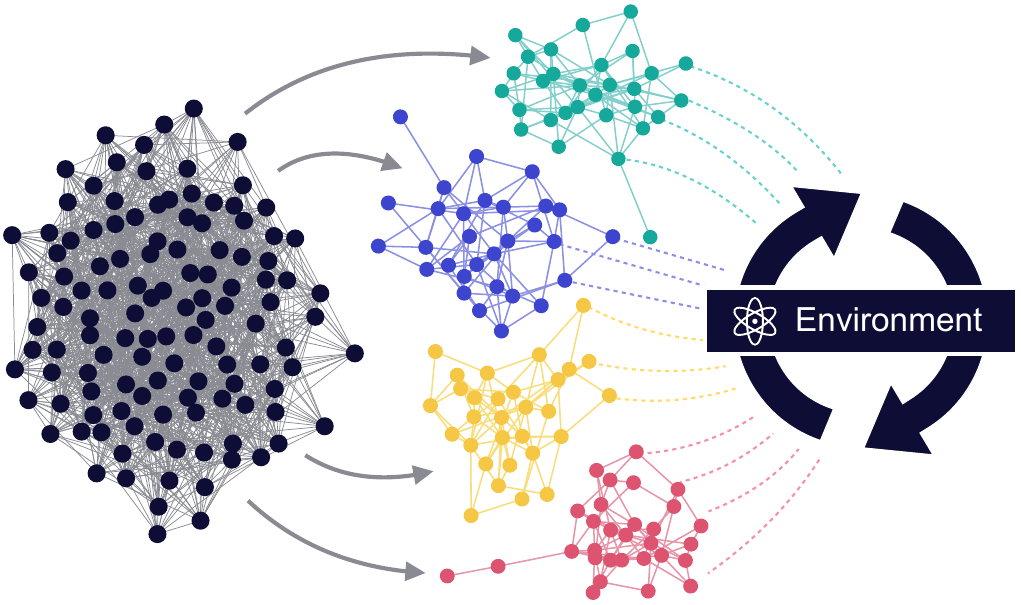}
    \caption{Schematic of the decomposition strategy. A problem is partitioned into $K=4$ subproblems of equal size. The interaction terms between subproblems are encoded in a mean-field environment, which is determined self-consistently via a variational quantum circuit.}
    \label{fig:introduction}
\end{figure}

Since optimization problems can be formulated as Ising Hamiltonians~\cite{lucas2014}, we draw upon ideas from many-body physics. In that context, a mean-field treatment addresses the intractability of some interacting systems by replacing specific interaction terms with their scalar expectation values~\cite{PhysRevB.95.020404,PhysRevB.94.144403,PhysRevB.102.195145,PhysRevX.13.011039,PhysRevB.111.125141}. This approximation effectively decouples complex many-body interactions into lower-dimensional subsystems embedded in an effective field---conceptually forming a shared environment (Fig.~\ref{fig:introduction}). This environment is determined through an iterative, self-consistent process where the field and system state mutually stabilize. Adapting this framework, we introduce a self-consistent mean-field quantum approximate optimization algorithm. By embedding inter-subproblem couplings into this environment, we solve manageable subproblems independently without neglecting their mutual influence, allowing them to be addressed individually rather than solving the entire original problem at once. This approach unlocks problem scales that would otherwise require prohibitive qubit and gate counts.

The remainder of the paper is organized as follows. First, we introduce the algorithm and analyze the self-consistency of the generated environment. Next, we benchmark performance against extensive numerical simulations on Sherrington-Kirkpatrick (SK) spin glasses~\cite{PhysRevLett.35.1792}---Ising models characterized by all-to-all two-body interactions with normally distributed strengths. We examine the impact of both the degree of decomposition and the underlying circuit depth, demonstrating that at shallow depths, the algorithm achieves parity with the standard QAOA on the full problem. Finally, we experimentally validate the self-consistent mean-field approach on a weighted maximum clique problem relevant to molecular docking for drug discovery~\cite{Kuhl1984}. Although this problem involves hundreds of degrees of freedom and thousands of interaction terms---exceeding the capacity of the standard QAOA on current hardware---our decomposition strategy enables the solution of such a real-world instance at scale.

\section{Self-Consistent Mean-Field Quantum Approximate Optimization}
\label{sec:method}

We are interested in finding the ground state of an Ising Hamiltonian of the form:
\begin{equation}
    \hat{C}=\sum_{i=1}^N\hat{Z}_i\left(h_i + \sum_{j=i+1}^N\mathsf{W}_{ij}\hat{Z}_j\right),
    \label{eq:obj_function}
\end{equation}
where $\hat{Z}_i$ is the Pauli-$Z$ operator acting on the $i$-th degree of freedom such that $\hat{Z}_i\vert0\rangle_i=\vert0\rangle_i$ and $\hat{Z}_i\vert1\rangle_i=-\vert1\rangle_i$. The parameters $\boldsymbol{h}\in\mathbb{R}^N$ and $\mathsf{W}\in\mathbb{R}^{N\times N}$ encode the problem, with symmetric couplings $\mathsf{W}_{ij}=\mathsf{W}_{ji}$ and null diagonal terms $\mathsf{W}_{ii}=0$. Without loss of generality, we assume that $\boldsymbol{h}\neq\boldsymbol{0}$~\footnote{The condition $\boldsymbol{h}\neq\boldsymbol{0}$ implies that Eq.~\eqref{eq:obj_function} does not possess a global $\mathbb{Z}_2$ spin-flip symmetry; i.e., the transformation $\hat{Z}_i\to -\hat{Z}_i$ for all $i$ does not leave the Hamiltonian invariant. If $\boldsymbol{h}=\boldsymbol{0}$, one can substitute one of the Pauli-$Z$ operators with one of its eigenvalues (e.g., $\hat{Z}_N\to\pm 1$). This recasts the problem over $N-1$ degrees of freedom with effective fields $h_i=\pm\mathsf{W}_{iN}$ for $i=1,2,\ldots, N-1$, leaving the Hamiltonian's spectrum invariant.}. 

When the scale of Eq.\eqref{eq:obj_function} renders it computationally intractable, a common approach is to decompose the problem into independent subproblems. For instance, some frameworks treat subproblems as strictly independent and focus on the ``stitching'' process, using either quantum optimization~\cite{zhou2023qaoa} or classical algorithms~\cite{vcepaite2025quantum} to assemble a global solution. Others introduce coupling directly into the architecture: Ref.~\cite{tomesh2023divide} utilizes partial mixers to link subproblems, while Refs.~\cite{tan2021qubit,Sundar2024} employ a qubit ``label'' register to encode all subproblems within a single quantum state. Recent hybrid efforts have further expanded this space, optimizing subproblem contributions against the full problem context~\cite{Liu2022} or leveraging tensor networks to manage inter-subproblem correlations~\cite{Huang2026}.

Here, we describe a decomposition strategy where we decompose a problem into subproblems, which we couple via a self-consistent mean-field approximation. Specifically, we factorize the Hamiltonian into $K$ independent Hilbert spaces, each comprising disjoint sets $\{S_1, S_2,\ldots S_K\}$ of $N/K$ degrees of freedom:
\begin{equation}
    \hat{\tilde{C}}=\sum_{n=1}^K\hat{C}_n(\boldsymbol{e}),\quad\hat{C}_n=\sum_{i\in S_n}\hat{Z}_i\left(\tilde{h}_i + \sum_{j\in S_n> i}\mathsf{W}_{ij}\hat{Z}_j\right),
    \label{eq:factorized_objective_function}
\end{equation}
where $\tilde{h}_i=h_i+\sum_{j\notin S_n> i} \mathsf{W}_{ij} e_j$ with $\boldsymbol{e}\in\mathbb{R}^N$ an environment capturing removed interactions between degrees of freedom in different Hilbert spaces. Taking $\boldsymbol{e}=\boldsymbol{0}$ corresponds to solving uncorrelated subproblems $\hat{C}_n$, as is done in, e.g., Refs.~\cite{zhou2023qaoa, vcepaite2025quantum} before stitching their respective solution together. The choice of environment $\boldsymbol{e}$ is critical for making the factorization as exact as possible with respect to finding the ground state of $\hat{C}$ through $\hat{\tilde{C}}$. For instance, taking $e_i=\langle\boldsymbol{x^*}\vert\hat{Z}_i\vert\boldsymbol{x^*}\rangle=\pm 1$ where $\vert\boldsymbol{x^*}\rangle$ is the unique ground state of the $\hat{C}$, means that $\vert\boldsymbol{x^*}\rangle$ is also the ground state of $\hat{\tilde{C}}$.

Inspired by Refs.~\cite{PhysRevB.95.020404,PhysRevB.94.144403,PhysRevB.102.195145,PhysRevX.13.011039,PhysRevB.111.125141}, we propose to build an environment $\boldsymbol{e}$ self-consistently via a hybrid classical-quantum algorithm such as the QAOA~\cite{Farhi2014}. For a Hamiltonian $\hat{H}$, the QAOA generates the unitary:
\begin{equation}
    \hat{U}=\prod_{k=1}^pe^{-i\beta_k\sum_{j\in S}\hat{X}_j}e^{-i\gamma_k\hat{H}},
    \label{eq:qaoa}
\end{equation}
where the Hamiltonian is defined over degrees of freedom $S$ and $\hat{X}_i$ is the Pauli-$X$ operator acting on the $i$-th degree of freedom. The unitary is parameterized by $\{\boldsymbol{\gamma},\boldsymbol{\beta}\}\in\mathbb{R}^{2p}$. The QAOA applied to the factorized Hamiltonian of Eq.~\eqref{eq:factorized_objective_function} yields a factorized unitary $\hat{\tilde{U}}=\bigotimes_{n=1}^K\hat{U}_n$. A similarly factorized quantum state can be built from it:
\begin{equation}
    \vert\Psi\rangle=\bigotimes_{n=1}^K\vert\phi_n\rangle,\quad\vert\phi_n\rangle=\hat{U}_n\vert+\rangle^{\otimes\vert S_n\vert},
    \label{eq:factorized_qaoa_state}
\end{equation}
where $\vert+\rangle_i=(\vert0\rangle_i+\vert1\rangle_i)/\sqrt{2}$ is motivated by the QAOA structure and its connection to the adiabatic theorem as $p\to+\infty$~\cite{Farhi2014}.

We proceed as follows to build the environment $\boldsymbol{e}$. The procedure begins with an initial guess $\boldsymbol{e}^{(\ell=0)}$ at step $\ell=0$. At step $\ell>0$, a subproblem $\hat{C}_n$ is selected at random and the corresponding state $\vert\phi_n\rangle$ computed with the QAOA. Then, we update the environment entries as $\boldsymbol{e}^{(\ell)}_{i}=\langle\Psi\vert\hat{Z}_i\vert\Psi\rangle=\langle\hat{Z}_i\rangle\in[-1,+1]$---only the entries corresponding to degrees of freedom of $\hat{C}_n$ are affected. This process repeats until self-consistency is achieved, as defined by a convergence threshold $\varepsilon$:
\begin{equation}
    \left\vert 1-\frac{\boldsymbol{e}^{(\ell)}}{\boldsymbol{e}^{(\ell+1)}}\right\vert<\varepsilon_e,~~\left\vert 1-\frac{\langle\hat{C}\rangle^{(\ell)}}{\langle\hat{C}\rangle^{(\ell+1)}}\right\vert<\varepsilon_c,
    \label{eq:convergence}
\end{equation}

\begin{figure*}[!t]
    \centering
    \includegraphics[width=1\textwidth]{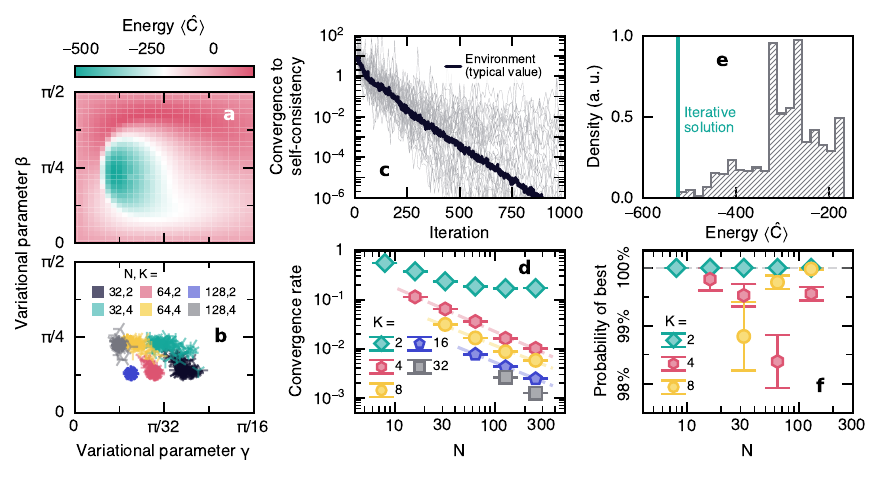}
    \caption{Analysis of self-consistency and solution landscapes. (a) Energy landscape for an $N=128$ SK problem instance decomposed into $K=4$ subproblems, plotted after environmental self-consistency is achieved for fixed QAOA parameters $\gamma$ and $\beta$. (b) Distribution of optimal QAOA parameters ($\gamma,\beta$) for hundreds of randomly generated SK instances with fixed $N$ and $K$ values. (c) Convergence to self-consistency for the environment, as defined in Eq.~\eqref{eq:convergence} for $N=128$ SK instances ($K=4$) using the $p=1$ QAOA, plotted against the number of iterations. (d) Convergence rate of the environment for typical SK problem instances at fixed, near-optimal parameters. (e) Distribution of energy obtained by solving the system of nonlinear equations (Eq.~\eqref{eq:sys_nonlin_eq}) for a single $N=128, K=4$ instance, using random initial guesses. The vertical line indicates the energy found by our iterative method starting from a null environment. (f) Probability that the iterative method (initialized with a null environment) converges to the lowest energy configuration found across random trials, averaged over SK instances with fixed parameters.}
    \label{fig:self_consistency}
\end{figure*}

Importantly, we utilize a common set of $2p$ variational QAOA parameters $\{\boldsymbol{\gamma},\boldsymbol{\beta}\}$ across all subproblems, holding them constant until environment self-consistency is attained. Effectively, this establishes a nested structure: within the broader variational optimization loop, the environment is converged to self-consistency for each specific trial set of variational parameters.

For a given subproblem $\hat{C}_n$, the above environment choice is equivalent to the following approximation in the original Hamiltonian of Eq.~\eqref{eq:obj_function}:
\begin{equation}
    \hat{Z}_{i\in S_n}\hat{Z}_{j\notin S_n}\approx\hat{Z}_{i\in S_n}\bigl\langle\hat{Z}_{j\notin S_n}\bigr\rangle,
    \label{eq:mf_approx}
\end{equation}
This method mirrors the mean-field approach used in quantum many-body physics, where operators acting on certain degrees of freedom are replaced by their scalar averages to simplify the underlying Hamiltonian~\cite{PhysRevB.95.020404,PhysRevB.94.144403,PhysRevB.102.195145,PhysRevX.13.011039,PhysRevB.111.125141}.

\section{Results}

\subsection{Self-consistency}

We begin by investigating the convergence of the environment toward self-consistency. We consider an ensemble of paradigmatic SK spin glasses~\cite{PhysRevLett.35.1792} with $N$ degrees of freedom. In Eq.~\eqref{eq:obj_function}, the couplings $\mathsf{W}_{ij}$ are normally distributed and the fields are set to $\boldsymbol{h}=\boldsymbol{0}$ (prior to the explicit breaking of the $\mathbb{Z}_{2}$ symmetry).

We randomly partition each instance into $K$ subproblems $\{S_1, \dots, S_K\}$ of approximately equal size $\vert S_{n=1,\ldots K}\vert\approx N/K$. Unless otherwise specified, we employ a null vector $\boldsymbol{e}^{(\ell=0)}=\boldsymbol{0}$ as the initial environment guess. We utilize the QAOA with $p=1$ layer, defined by two variational parameters $\gamma$ and $\beta$~\cite{Farhi2014}. Observables such as $\langle\hat{Z}_i\rangle$ and $\langle\hat{Z}_i\hat{Z}_j\rangle$ are computed analytically via back-propagation~\cite{PhysRevA.109.012429}; these values are subsequently used to compute the total energy obtained by solving the independent subproblems within their self-consistent environment.

\subsubsection{Variational optimization and self-consistency}

First, we examine the energy landscape for the case of an $N=128$ instance decomposed into $K=4$ subproblems. We achieve self-consistency of the environment by setting $\varepsilon=10^{-4}$ in Eq.~\eqref{eq:convergence}. Figure~\ref{fig:self_consistency}a displays the resulting energy landscape obtained by combining the $K$ subproblems with their self-consistent environment. The landscape exhibits a distinct structure, similar to that obtained with standard QAOA. The landscape is smooth, which allows one to carry out the variational search.

We repeat a similar calculation as the above, for an ensemble of randomly generated instances, for different $(N, K)$ values. For each problem instance, we numerically optimize the variational parameters using the Nelder-Mead algorithm~\cite{Gao2012}. As shown in Fig.~\ref{fig:self_consistency}b, we observe parameter concentration. Parameter concentration is known to occur in the QAOA~\cite{Brandao2018,Farhi2022}, with the optimal parameters for the SK model at $p=1$ being $\beta=\pi/8$ and $\gamma= 1/2\sqrt{N}$. Fig.~\ref{fig:self_consistency}b demonstrates that it also occurs within the self-consistent mean-field framework. We find that the phase separator parameter $\gamma$ scales as $N^{-1/2}$ for all $K$, mirroring the behavior of the standard QAOA ($K=1$) for SK models~\cite{Farhi2022}. The mixer parameter $\beta$ is independent of $N$. Both parameters depend on the number of subproblems $K$. Moreover, as the problem size $N$ increases, the spread of optimal parameters across instances diminishes. This implies that parameter transfer techniques developed for the QAOA~\cite{8916288,PhysRevX.10.021067,PhysRevA.103.042612,Galda2021,Farhi2022,Galda2023,10.1145/3584706,shaydulin2024evidence} can be leveraged to mitigate variational optimization overhead.

The dependence of $\beta$ on $K$ can be understood from the contribution of the environment to the energy. At $K=1$, the energy (Eq.~\eqref{eq:factorized_objective_function}) is extremized through terms like $\mathsf{W}_{ij}\langle\hat{Z}_i\hat{Z}_j\rangle$. Ref.~\cite{Farhi2022} showed that these correlations are extremized for $p=1$ at $\beta=\pi/8$. Conversely, as $K\to N$, all the terms in the energy are determined by one-body expectations via $h_i\langle\hat{Z}_i\rangle$ and $\mathsf{W}_{ij}\langle\hat{Z}_i\rangle\langle\hat{Z}_j\rangle$. A similar calculation as Ref.~\cite{Farhi2022} will show that these one-body expectations are extremized at $\beta=\pi/4$. This calculation is discussed further in Section~\ref{sec:performance}.

\subsubsection{Convergence}

We investigate the convergence of the environment to self-consistency using fixed, optimized variational parameters. Figure~\ref{fig:self_consistency}c displays data for $N=128$ SK problem instances decomposed into $K=4$ subproblems as a function of the number of iteration steps.

Observing that the convergence of individual instances spans orders of magnitude, we analyze the typical value defined as $\textrm{typ}(x)=\exp\mathbb{E}\left[\log x\right]$ where $\mathbb{E}[x]$ denotes the mean. We find that typical environment convergence is asymptotically exponential with the number of iterations $\ell$:
\begin{equation}
    \mathbb{E}\left[\log\bigl\vert 1-{\boldsymbol{e}^{(\ell)}\bigr/}{\boldsymbol{e}^{(\ell+1)}}\bigr\vert\right]\sim -\ell f\bigl(N,K\bigr),
\end{equation}
where the convergence rate $f(N,K)$ is estimated via least-squares fitting (Fig.~\ref{fig:self_consistency}d). For a fixed number of subproblems $K>2$, the convergence rate scales polynomially with $N$, exhibiting a sublinear exponent ($\approx -0.85$) that appears roughly independent of $K$. However, the $K=2$ case exhibits distinct behavior, with the convergence rate saturating with problem size $N$. This regime is unique because, for SK instances with $K=2$, exactly half of the interaction terms are treated exactly within the subproblems, while the other half are mediated via the mean-field environment. In the general asymptotic limit ($N\rightarrow+\infty$), a density of $1-1/K$ of the terms are treated through the environment.

\subsubsection{Chaotic behavior and self-consistent solutions}
\label{sec:sys_nonlineq}

It is interesting to ask whether the self-consistent solutions are unique solutions, or whether they at least yield the lowest-energy solutions among a multitude of possible self-consistent solutions. We begin by writing the explicit analytical expression for the expectation value $\langle\hat{Z}_{i}\rangle$, limiting to $p=1$ for now. From Ref.~\cite{PhysRevA.109.012429}, we have that:
\begin{multline}
    0=\bigl\langle\hat{Z}_{i\in S_n}\bigr\rangle+\sin\left(2\beta\right)\sin\left[2\gamma\left(h_i+\sum\nolimits_{j\notin S_n}\mathsf{W}_{ij}e_j\right)\right]\\
    \times\prod\nolimits_{j\in S_n}\cos\left(2\gamma\mathsf{W}_{ij}\right),
    \label{eq:sys_nonlin_eq}
\end{multline}
where self-consistency with $e_j\equiv\langle\hat{Z}_j\rangle$ implies a system of $N$ nonlinear equations. Solving this system for the variables $\langle\hat{Z}_i\rangle$ leads to a self-consistent solution. Crucially, the iterative method developed in Section~\ref{sec:method} solves these equations implicitly without requiring their analytical derivation, thus facilitating extension to $p>1$ and non-QAOA circuits.

Equation~\eqref{eq:sys_nonlin_eq} offers an alternative perspective on the approach: such systems are generically difficult to solve, often admitting multiple solutions and exhibiting sensitivity to initial seeds and solver choice. We illustrate this behavior by solving Eq.~\eqref{eq:sys_nonlin_eq} for SK instances using \texttt{MINPACK}'s \texttt{hybrid} method~\cite{MoreGarbowHillstrom1980}. Fixing the variational parameters to their optimized values, we test thousands of initial guesses drawn uniformly at random as $\langle\hat{Z}_i\rangle\in[-1,+1]\;\forall i$.

We find multiple distinct solutions satisfying the self-consistency condition within numerical precision. This is exemplified in Fig.~\ref{fig:self_consistency}e, which displays the distribution of energy $\langle\hat{C}\rangle$ for the self-consistent environments of a single instance ($N=128, K=4$). Notably, the energy obtained via our iterative method (starting from a null environment, $\boldsymbol{e}=\boldsymbol{0}$) corresponds to the minimum energy solution. To test the generality of this result, we calculate the probability that the iterative method finds an energy within $0.1$ of the global minimum sampled. As shown in Fig.~\ref{fig:self_consistency}f, this probability exceeds $98\%$ across various $N$ and $K$ values, highlighting the remarkable robustness and stability of the proposed method despite the underlying chaotic landscape. At the same time, there may exist other variational parameters where other self-consistent environments would lead to a better energy.

\subsection{Performance}
\label{sec:performance}

We evaluate the performance of the proposed self-consistent mean-field QAOA by measuring the expected value of the energy $\langle\hat{C}\rangle$ over the QAOA subproblem quantum states. We study SK problem instances with $N$ degrees of freedom and decomposed into $K$ subproblems of approximately equal size. In the asymptotic limit $N\to+\infty$, the absolute ground state energy of SK problem instances is known to go as $\sim P^*N^{3/2}$ with $P^*\simeq -0.7631\ldots$ the Parisi constant~\cite{PhysRevLett.50.1946}.

\subsubsection{Number of subproblems}

\begin{figure}[!t]
    \centering
    \includegraphics[width=1\columnwidth]{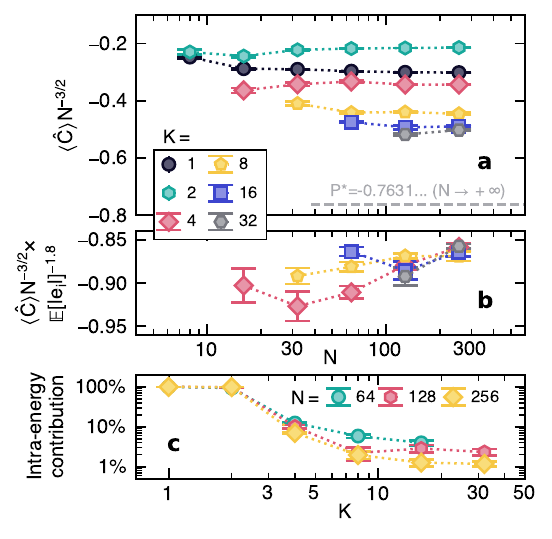}
    \caption{Performance scaling on SK instances with $N$ degrees of freedom decomposed into $K$ subproblems. (a) Energy density obtained via the self-consistent mean-field QAOA with $p=1$ layer. The $K=1$ case corresponds to the standard QAOA results~\cite{Farhi2022}. $P^*$ denotes the exact ground state energy density for infinite-size SK instances (Parisi constant)~\cite{PhysRevLett.50.1946}. (b) Energy density for $K>2$, rescaled by a power of the average environment magnitude. (c) Relative contribution of intra-subproblem terms to the total energy. Data points represent averages over tens to hundreds of random instances; error bars indicate the standard error of the mean. Dotted lines are guides to the eye.}
    \label{fig:performance_p=1}
\end{figure}

We focus on the $p=1$ case for the self-consistent mean-field QAOA. In the absence of an environment ($\boldsymbol{e}=\boldsymbol{0}$), the subproblems behave as independent SK instances; consequently, one anticipates the energy density $\langle\hat{C}\rangle N^{-3/2}\approx -1/\sqrt{4Ke}$~\footnote{Under depth-$p$ QAOA, the average energy for SK problems (Eq.~\eqref{eq:obj_function}) scales asymptotically with system size $N$ as $\langle\hat{C}\rangle_p\sim \alpha_p N^{3/2}$, where $\alpha_p\sim O(1)$ is the energy density~\cite{Farhi2022}. Upon decomposition, the problem splits into $K$ independent subproblems, each effectively an SK instance with $N/K$ variables. Consequently, the combined energy of the factorized system (Eq.~\eqref{eq:factorized_objective_function}) scales as $\langle\hat{\tilde{C}}\rangle\sim K\alpha_p(N/K)^{3/2}$. This yields a relative energy scaling of $\langle\hat{\tilde{C}}\rangle\sim \langle\hat{C}\rangle_p/\sqrt{K}$. For the specific case of $p=1$, the density is given by $\alpha_{1}=-1/\sqrt{4e}$~\cite{Farhi2022}.}. Figure~\ref{fig:performance_p=1}a displays the energy density against $N$ for various values of $K$, averaged over tens to hundreds of random instances, where the environment was computed to self-consistency for each instance. The $K=1$ case corresponds to the standard QAOA~\cite{Farhi2022}. The $K=2$ case has a smaller cost magnitude numerically consistent with a factor $1/\sqrt{2}$ times the cost for $K=1$, despite the presence of a nonzero environment. Intuitively, increasing the decomposition granularity ($K$) reduces the quantum correlations explicitly handled within subproblems, which one might expect to degrade solution quality. However, we observe the opposite trend for $K>2$: increasing $K$ lowers the energy, improving performance.

We attribute this counter-intuitive trend to the increasing average magnitude of the environment entries, $\mathbb{E}[\vert e_i\vert]$. As shown in Fig.~\ref{fig:performance_p=1}b, we can eliminate this $K$ dependence by empirically rescaling the energy density by $\mathbb{E}[\vert e_i\vert]^{-1.8}$, though the limited range of accessible $K$ values prevents us from extracting a definitive phenomenological scaling. Nevertheless, we gain insight from the $K\to N$ limit, where contributions to the energy $\langle\hat{\tilde{C}}\rangle$ arise exclusively from one-body expectation values (via $h_i\langle\hat{Z}_i\rangle$ and $\mathsf{W}_{ij}\langle\hat{Z}_i\rangle\langle\hat{Z}_j\rangle$); effectively, the energy is determined by the environment $\boldsymbol{e}$ itself. Consequently, energy extremization requires $\vert e_i\vert$ increasing with $K$. Since Eq.~\eqref{eq:sys_nonlin_eq} establishes that $\vert e_i\vert\propto\sin2\beta$, this requirement aligns with the observation that $\beta\rightarrow\pi/4$ as $K$ increases (see Fig.~\ref{fig:self_consistency}b).

In the thermodynamic limit, a fraction $1/K$ of the terms are treated exactly within subproblems, while the rest are mediated by the environment. The critical value is $K=2$ and can help understand the observed behavior. As illustrated in Fig.~\ref{fig:performance_p=1}c, the total energy is dominated by intra-subproblem contributions for $K\leq 2$, whereas inter-subproblem terms (mediated by the environment) dominate for $K>2$.

While the limit $K\to N$ is theoretically accessible, we observe in Fig.~\ref{fig:performance_p=1}a that the energy density saturates at approximately $\langle\hat{C}\rangle N^{-3/2}\approx -0.5$, which remains above the Parisi constant $P^*$. This value is analogous to what one would obtain with a classical greedy solver~\cite{Dupont2023} or the standard QAOA at $p=4$~\cite{Farhi2022}, but higher compared to that of a relax-and-round solver~\cite{Aizenman1987,PhysRevA.109.012429}. Also, increasing $N$ and $K$ increases the difficulty of achieving self-consistency (see Figs.~\ref{fig:self_consistency}c and~\ref{fig:self_consistency}d), necessitating a careful balance between decomposition granularity and computational overhead. Nevertheless, for intermediate problem sizes, the proposed framework provides a pathway to investigate instances that are otherwise inaccessible to the standard QAOA due to prohibitive qubit or gate requirements.

\subsubsection{Number of variational parameters}

\begin{figure}[!t]
    \centering
    \includegraphics[width=1\columnwidth]{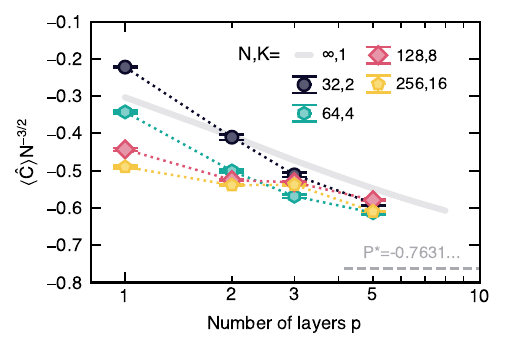}
    \caption{Impact of circuit depth. Average energy density for SK instances ($N/K \approx 16$) as a function of QAOA layers $p$, using the self-consistent mean-field approach. The limit $N \rightarrow +\infty$ with $K=1$ corresponds to the standard QAOA~\cite{Farhi2022}. Data points represent averages over tens to hundreds of random instances; error bars indicate the standard error of the mean. Dotted lines are guides to the eye.}
    \label{fig:performance_nlayers}
\end{figure}

We examine SK problem instances with a fixed ratio $N/K\approx 16$ and circuit depth $p \geq 1$, employing the Nelder-Mead algorithm to optimize the $2p$ variational QAOA parameters. Consistent with our previous methodology, these parameters are shared across all subproblems, and environment self-consistency is established for each set of trial parameters. Figure~\ref{fig:performance_nlayers} displays the average expected energy density as a function of the number of QAOA layers $p$. We observe that the performance gap between different decomposition counts $K$---prominent at $p=1$---diminishes as $p$ increases. Furthermore, performance improves monotonically within the considered regime ($p\leq 5$).

Quantum optimization algorithms grounded in the adiabatic theorem generally exhibit performance improvements with increasing circuit depth. For instance, the QAOA is known to converge asymptotically to the ground state as $p\to+\infty$~\cite{Farhi2014,Wurtz2022}. Recent numerical evidence suggests that the QAOA yields a $(1-\delta)$ approximation to the ground state energy of SK problem instances with circuit depth $p\sim1/\delta^{1.13}$ in the average case~\cite{Boulebnane2025}.

\begin{figure*}[!t]
    \centering
    \includegraphics[width=1\textwidth]{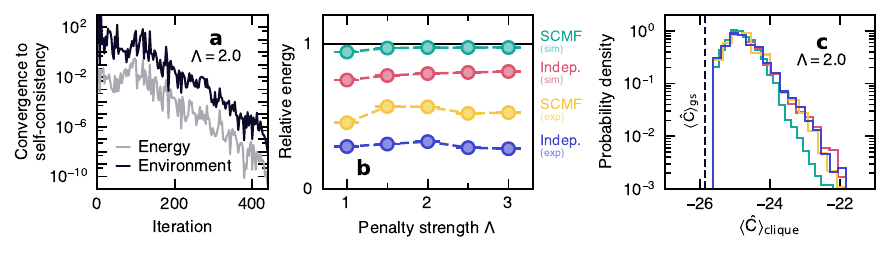}
    \caption{Application to molecular docking, formulated as a weighted maximum clique problem. (a) Convergence of the environment and energy towards self-consistency (Eq.~\eqref{eq:convergence}) as a function of iterations, using $p=1$ QAOA with fixed variational parameters and $\Lambda=2$. (b) Relative energy versus penalty strength $\Lambda$, with random sampling and standard $p=1$ QAOA serving as baselines. Data points represent averages over $10^4$ samples; error bars indicate the standard error of the mean and dotted lines are guides to the eye. (c) Energy probability density for $10^4$ classically post-processed samples where the clique constraint is enforced. The color scheme matches panel (b), and the absolute ground state energy is indicated by $\langle\hat{C}\rangle_\textrm{gs}\simeq -25.86$.}
    \label{fig:molecular_docking}
\end{figure*}

The self-consistent mean-field QAOA recovers the standard QAOA in the limit where the number of inter-subproblem terms vanishes, or equivalently, as $K\to 1$. However, taking the $K\to 1$ limit negates the computational advantages of the decomposition approach. Moreover, the limits $p\to+\infty$ and $K\to 1$ do not necessarily commute; which means that $p\to+\infty$ at a fixed $K>1$ does not guarantee the optimal solution. Consequently, we anticipate that increasing $p$ further would result in energy density saturation $\langle\hat{C}\rangle N^{-3/2}>P^*$.

Nevertheless, for the shallow circuits considered ($p\leq 5$), the performance is comparable to the standard QAOA~\cite{Farhi2022}. This positions the framework as a potent method for embedding shallow quantum optimization circuits into problems that would otherwise exceed the qubit and gate capacities of current hardware.

\subsection{Application to a weighted maximum clique problem}
\label{sec:exp_maxclique}

\subsubsection{Problem and implementation}

We extend our analysis beyond SK spin glasses to a weighted maximum clique problem, formulated as the ground state of a Hamiltonian~\cite{lucas2014}:
\begin{equation}
    \hat{C}(\Lambda)=\sum_{i=1}^N\hat{n}_i\left[-h_i+\Lambda\sum_{j=i+1}^N\bigl(1-\mathsf{W}_{ij}\bigr)\hat{n}_{j}\right],
    \label{eq:maxclique_hamiltonian}
\end{equation}
where $\Lambda$ is a user-defined parameter and $\hat{n}_i$ is the occupation operator acting on the $i$-th degree of freedom such that $\hat{n}_i\vert0\rangle_i=0$ and $\hat{n}_i\vert1\rangle_i=\vert1\rangle_i$. The Hamiltonian parameters $\boldsymbol{h}\in\mathbb{R}^N$ and $\mathsf{W}\in\{0,1\}^{N\times N}$ ($\mathsf{W}_{ij}=\mathsf{W}_{ji}$ and $\mathsf{W}_{ii}=0$) are derived from a real-world molecular docking dataset relevant to drug discovery (see Appendix~\ref{app:moldock_to_maxclique})~\cite{Kuhl1984}. The precise problem instance comprises $N=252$ degrees of freedom and $4,257$ nonzero two-body interaction terms, making it significantly sparser than the SK models considered previously. Although this Hamiltonian can be recast as an Ising Hamiltonian in the form of Eq.~\eqref{eq:obj_function} with Pauli-$Z$ operators, we work with the binary version and adapt the self-consistent mean-field framework accordingly. We choose $e_i=\langle\hat{n}_i\rangle\in[0,1]$ for the environment with $\boldsymbol{e}^{(\ell=0)}=\boldsymbol{1/2}$ as an initial guess.

We focus on the $p=1$ QAOA. Implementing the full problem on a superconducting platform with limited connectivity incurs substantial compilation overhead. Specifically, compiling an optimized $p=1$ circuit for this instance onto a linear topology of $N=252$ qubits requires approximately $8.5 \times 10^4$ two-qubit gates ($\texttt{CZ}$ or $\texttt{ISWAP}$)~\cite{MATSUO20232022EAP1159}, a requirement that exceeds current hardware capabilities. To mitigate this, we randomly decompose the problem into $K=12$ subproblems, each with $N/K=21$ degrees of freedom. This reduces the resource requirements to linear chains of $21$ qubits with a number of two-qubit gates ranging from $97$ to $227$---a reduction of more than two orders of magnitude.

The experiments were conducted on the Rigetti Ankaa-3 superconducting quantum processor. The system operates with average fidelities of approximately $99.0\%$ for two-qubit $\texttt{ISWAP}$ gates, $99.8\%$ for single-qubit $\texttt{RX}$ gates, and $97.9\%$ for readout.

\subsubsection{Results}

Setting $\Lambda=2$ and using a depth-$1$ QAOA, we numerically optimize the variational parameters via the Nelder-Mead algorithm. Fig.~\ref{fig:molecular_docking}a illustrates the convergence of both the environment and the energy toward self-consistency (Eq.~\eqref{eq:convergence}). Both metrics exhibit exponential decay with the number of iterations, consistent with our previous observations for the SK model. Upon achieving self-consistency, we sample $10^4$ solutions for each subproblem. To construct global candidate solutions, we independently draw samples from the solution pools of each subproblem and concatenate them.

Fig.~\ref{fig:molecular_docking}b shows the relative energy as a function of penalty strength $\Lambda$. For reference, random solutions and the standard $p=1$ QAOA are scaled to $0$ and $1$, respectively. We compare the self-consistent mean-field approach against independent QAOAs solving the subproblems. The self-consistent mean-field yields superior solution quality compared to the independent approach, performing comparably to the standard QAOA despite the resource reduction achieved through decomposition. While hardware noise degrades the absolute quality of the experimental runs, the performance hierarchy---where self-consistent mean-field outperforms independent QAOAs---remains intact.

Unlike the SK analysis, which relies on expectation values, the maximum clique problem requires analyzing individual solutions. Because the Hamiltonian employs soft penalties, raw samples from an emulation of or experiment on the quantum processor may not represent valid cliques. To enforce validity, we apply a classical greedy post-processing step to each solution (Appendix~\ref{app:clique_postprocess}). Fig.~\ref{fig:molecular_docking}c displays the energy distribution of these post-processed solutions at $\Lambda=2$, spanning the range from $\langle0\ldots 0\vert\hat{C}\vert0\ldots 0\rangle=0$ to $\langle\hat{C}\rangle_\textrm{gs}\simeq -25.86$ (estimated via simulated annealing~\cite{kirkpatrick1983optimization}). We find that all methods (both simulation and experiment) can reach $\gtrsim 99\%$ of the ground state energy. While Fig.~\ref{fig:molecular_docking}b showed a distinct separation between methods, classical post-processing blurs this distinction. However, the tail of the simulated self-consistent mean-field distribution remains skewed toward higher-quality solutions.

Ultimately, constrained optimization problems present challenges for quantum optimization. The unique challenge for decomposition strategies is that even at large depth $p$, the global solution obtained by stitching together local solutions may not be optimal, and may not satisfy the constraints either. This necessitates heavy reliance on classical post-processing. This suggests that while decomposition effectively reduces the resource overhead for the quantum processor, standard implementations may be suboptimal for problems with hard, global constraints. Future work should focus on adapting this framework---perhaps by incorporating constraints directly into the self-consistency algorithm---or exploring alternative architectures specifically designed to preserve global validity during the recomposition process.

\subsubsection{Comparison with the existing literature}

Our work complements and extends recent efforts to apply quantum algorithms to molecular docking. While Ref.~\cite{doi:10.1126/sciadv.aax1950} pioneered the use of Gaussian boson sampling~\cite{hamilton2017gaussian} for this task---establishing the weighted maximum clique formulation we adopt---their approach relies on specialized photonic hardware distinct from the universal gate-model systems considered here. In the context of gate-based quantum computing, Refs.~\cite{PhysRevApplied.21.034036} and~\cite{Papalitsas2025} have investigated the application of the QAOA and the digitized-counterdiabatic QAOA~\cite{chandarana2022digitized,wurtz2022counterdiabaticity} to molecular docking problems.

These works could only access small-scale molecular docking instances with tens of variables. By contrast, our self-consistent mean-field strategy explicitly addresses the scale bottleneck via decomposition. This capability allows us to experimentally tackle a realistic instance with $252$ variables and offers a concrete pathway for investigating relevant molecular simulations with quantum algorithms.

\subsubsection{Effect of noise and environment scaling}

Under a fully depolarizing noise model, expectation values such as $\langle\hat{Z}_i\rangle$ or $\langle\hat{n}_i\rangle$ are damped by the circuit fidelity $F$. Consequently, the magnitude of the environment entries $\vert e_i\vert$ is effectively bounded by the fidelity, suppressing the feedback signal necessary for self-consistency. To counteract this, one can employ error mitigation techniques that rescale expectation values, such as zero-noise extrapolation~\cite{9259940}. Alternatively, we can introduce a scaling hyperparameter, $\boldsymbol{e}\leftarrow\eta\boldsymbol{e}$, to artificially restore the magnitude of the environment field. Notably, this parameter $\eta$ is versatile: even in a noiseless context, it can serve as a control knob to tune the environment strength relative to the local Hamiltonian.

\section{Conclusion}

In this work, we introduced a self-consistent mean-field quantum optimization algorithm designed to approximate the ground state of classical Ising Hamiltonians. We decomposed the global problem into subproblems, and mediated inter-subproblem couplings through a self-consistent environment generated via variational quantum circuits. We analyzed the convergence properties of this environment and benchmarked the algorithm's performance on SK spin glasses, finding that the method performs comparably to the standard QAOA at shallow circuit depths. Moreover, we experimentally validated the approach on a weighted maximum clique problem applied to molecular docking, successfully tackling a real-world instance with hundreds of variables---a scale exceeding the qubit and gate capacity of standard QAOA on contemporary hardware.

Our results raise several questions regarding the dynamics of the mean-field environment. We observed that convergence behavior depends on decomposition granularity, particularly distinguishing the $K=2$ regime from higher $K$. Additionally, there appear to be multiple stable solutions for the self-consistent environment, but our iterative method robustly converges to the environment corresponding to the minimal energy configuration with high probability. Developing a deeper understanding of this solution landscape would help to delineate the approach's limitations and optimize the trade-off between decomposition and computational overhead. Finally, this self-consistent mean-field framework is not inherently restricted to two-body Hamiltonians or the QAOA; future work should explore its application to alternative quantum optimization strategies.

Looking beyond optimization, the proposed approach holds promise for broader applications in quantum computing. The ability to reduce problem dimensionality while preserving essential correlations suggests utility in areas such as physical simulations~\cite{bauer2020quantum} and quantum machine learning~\cite{cong2019quantum,cerezo2022challenges}. For instance, by embedding large-scale datasets or complex quantum neural network architectures into decomposable sub-units interacting via a mean-field environment, one could tackle problems that are currently prohibitive for existing devices due to resource constraints. Such a framework thereby provides a platform to accelerate algorithm development and benchmarking of problems at scale on current quantum hardware.

\begin{acknowledgments}
    This work was partly supported by the U.S. Department of Energy, Office of Science, National Quantum Information Science Research Centers, Superconducting Quantum Materials and Systems Center (SQMS), under Contract No. 89243024CSC000002. Fermilab is operated by Fermi Forward Discovery Group, LLC under Contract No. 89243024CSC000002 with the U.S. Department of Energy, Office of Science, Office of High Energy Physics. This research used resources of the National Energy Research Scientific Computing Center, a DOE Office of Science User Facility supported by the Office of Science of the U.S. Department of Energy under Contract No. DE-AC02-05CH11231 using NERSC award ASCR-ERCAP0031818. The experiments were performed through Rigetti Computing Inc.'s Quantum Cloud Services QCS\textsuperscript{TM} on the Ankaa\textsuperscript{TM}-3 superconducting quantum processor developed, fabricated, and operated by Rigetti Computing Inc.
\end{acknowledgments}

\section*{Author contributions}

M.D. conceived and led the project. M.D. performed simulations, data collection, and data analyses. M.D. wrote the manuscript with input from B.S. and M.G. All coauthors contributed to the discussions leading to the completion of this project.

\section*{Competing interests}

M.D., B.S., and M.G. are, have been, or may in the future be participants in incentive stock plans at Rigetti Computing Inc. M.D. is inventor on a pending patent application related to this work (No.~63/998,278).

\section*{Data availability}

The molecular docking problem instance used in this work, and formatted as a graph, is publicly available at~\href{https://doi.org/10.5281/zenodo.18172244}{doi.org/10.5281/zenodo.18172244}~\cite{data_availability}.

\appendix

\section{Molecular docking as a weighted maximum clique problem}
\label{app:moldock_to_maxclique}

\begin{figure}[!t]
    \centering
    \includegraphics[width=0.8\columnwidth]{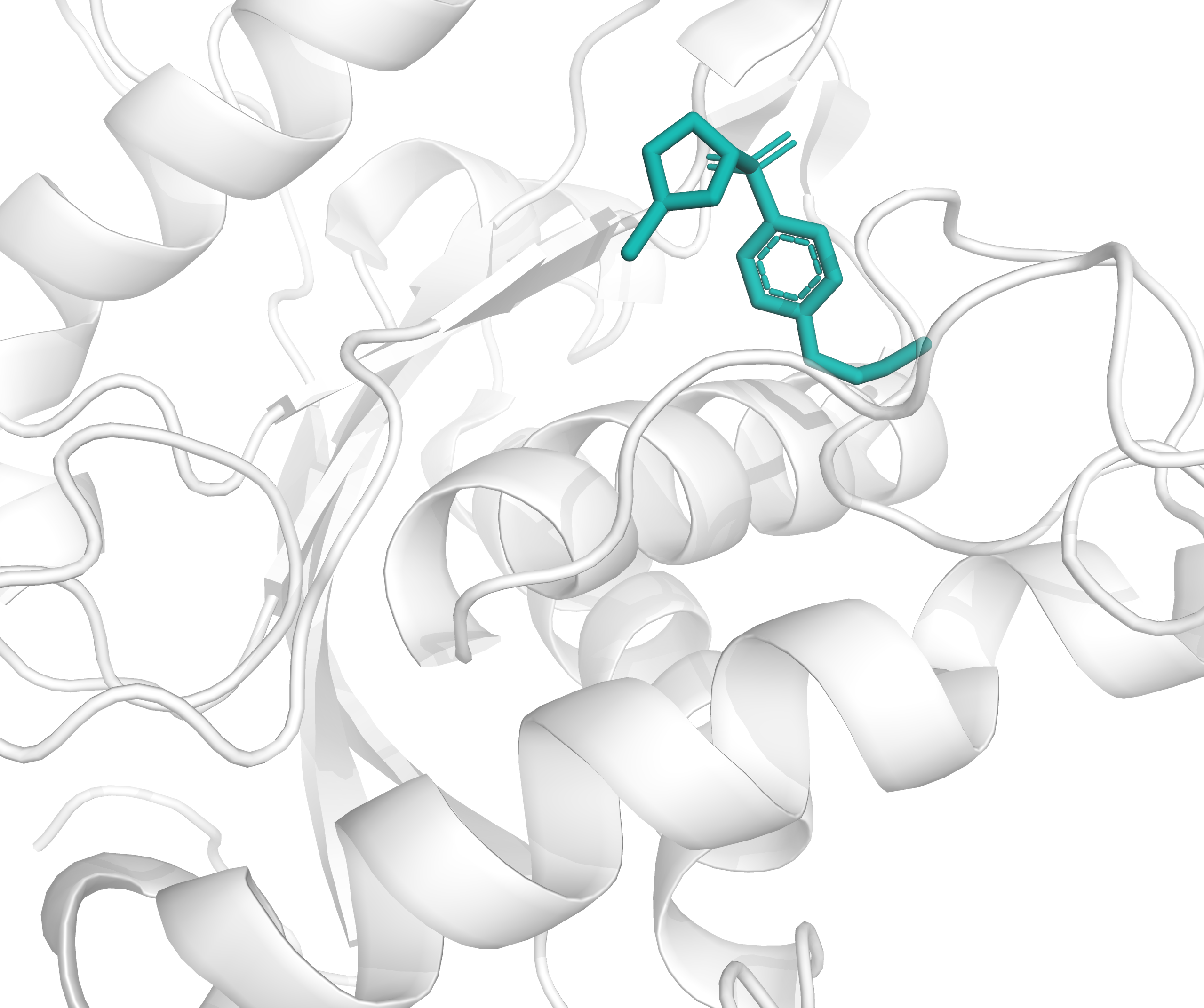}
    \caption{Visualization of the ligand docking problem, where we optimize the orientation of a molecule binding to a protein. This corresponds to the protein-ligand complex with PDB identifier~\texttt{2OI0} from the Protein Data Bank. This complex corresponds to a tumor necrosis factor-$\alpha$ converting enzyme bound to a thiol-containing aryl sulfonamide compound~\cite{GOVINDARAO20072250}.}
    \label{fig:molecular_docking_img}
\end{figure}

We follow Refs.~\cite{doi:10.1126/sciadv.aax1950,PhysRevApplied.21.034036,Papalitsas2025} to formulate the molecular docking problem discussed in Sec.~\ref{sec:exp_maxclique} as a weighted maximum clique problem in the form of Eq.~\eqref{eq:obj_function}~\cite{Kuhl1984}. 

\subsubsection{Input data and pre-processing}

We retrieve the protein-ligand complex with PDB identifier~\texttt{2OI0} from the Protein Data Bank. This complex corresponds to a tumor necrosis factor-$\alpha$ converting enzyme bound to a thiol-containing aryl sulfonamide compound~\cite{GOVINDARAO20072250} (see Fig.~\ref{fig:molecular_docking_img}). The input data is processed using~\texttt{ProDy}~\cite{10.1093/bioinformatics/btab187}, treating the protein and ligand as separate objects. Next, we extract $m=1,\dots,M$ interaction points on the ligand and $n=1,\dots,N$ interaction points on the protein. These points, derived from relevant chemical features using~\texttt{RDKit}~\cite{rdkit}, are known as pharmacophores. We consider six pharmacophore types: positive and negative charges, hydrogen-bond donors and acceptors, hydrophobic, and aromatic. The numerical interaction strengths $V_{t,t'}\geq 0$ between pharmacophore types $t$ and $t'$ are adopted from Ref.~\cite{doi:10.1126/sciadv.aax1950}. Additionally, we compute the Euclidean distances between all pairs of pharmacophores within the ligand $\mathsf{D}^{(\ell)}\in\mathbb{R}^{M\times M}$ and within the protein $\mathsf{D}^{(p)}\in\mathbb{R}^{N\times N}$.

\subsubsection{Problem construction}

The molecular docking problem aims to identify the optimal interaction zone between the protein and the ligand, subject to geometric constraints. We formulate this as a weighted maximum clique problem.

First, we define a binary variable $x_i\in\{0,1\}$ with $i=1,\ldots, I$ for each pair $i\equiv (m,n)$ consisting of a ligand pharmacophore $m$ and a protein pharmacophore $n$, such that the total number of variables is $I=M \times N$. Setting $x_i=1$ indicates that the corresponding pharmacophore pair is interacting. Each variable $i$ is associated with an interaction strength $V_{t,t'}$ based on the types $t$ and $t'$ of the pharmacophores involved, which we denote $h_i\geq 0$. The objective is to maximize the total interaction score:
\begin{equation}
    \max\sum_{i=1}^Ih_ix_i,
\end{equation}
subject to geometric compatibility constraints. Two pairs $i\equiv (m,n)$ and $i'\equiv (m',n')$ are deemed mutually compatible if the Euclidean distances between the ligand pharmacophores $(m, m')$ and protein pharmacophores $(n, n')$ match within a tolerance $\Delta_{mnm'n'}$:
\begin{equation}
    \left\vert\mathsf{D}_{mm'}^{(p)}-\mathsf{D}_{nn'}^{(\ell)}\right\vert\leq\Delta_{mnm'n'}.
\end{equation}
We encode this compatibility via the matrix $\mathsf{W}\in\{0,1\}^{I\times I}$, where $\mathsf{W}_{ii'}=1$ if pairs $i$ and $i'$ are compatible, and $0$ otherwise. The problem can then be stated as:
\begin{equation}
    \max\sum_{i=1}^Ih_ix_i\quad\textrm{s.t.}\quad \bigl(1 - \mathsf{W}_{ii'}\bigr)x_ix_{i'}=0~\forall i\neq i',
\end{equation}
which corresponds to finding the maximum weighted clique in the graph defined by $\mathsf{W}$. To map this to a ground state problem, we convert the maximization into a minimization and incorporate the hard constraints as soft penalties:
\begin{equation}
    C(\boldsymbol{x})=-\sum_{i=1}^Ih_ix_i+\Lambda\sum_{i=1}^I\sum_{i'=i+1}^I\bigl(1 - \mathsf{W}_{ii'}\bigr)x_ix_{i'},
    \label{eq:maxclique}
\end{equation}
where $\Lambda$ is the penalty strength. To ensure the ground state corresponds to a valid clique, one typically requires $\Lambda>2\max(h_i)$, although a lower bound may exist depending on the exact problem.

\subsubsection{Problem definition}

We make the resulting problem instance publicly available~\cite{data_availability}. After loading the chemical complex, we restrict the search space to ligand and protein pharmacophores located within a $5\,\text{\AA}$ radius of each other. We use a constant geometric tolerance $\Delta_{mnm'n'}\equiv\Delta=8.6\,\text{\AA}$ for all pairs. Considering only the six pharmacophore types with nonzero interaction potentials, we identify $7$ pharmacophores on the ligand and $36$ on the protein. This yields a problem of the form in Eq.~\eqref{eq:maxclique} with $I=7\times 36=252$ variables and $4,257$ nonzero quadratic terms.

\section{Classical post-processing for clique constraint}
\label{app:clique_postprocess}

The QAOA or self-consistent mean-field QAOA may sample solutions that are not valid cliques. To ensure that only valid solutions are considered, we perform a classical post-processing step on each sampled solution $\boldsymbol{x}$. We implement the greedy strategy of Ref.~\cite{doi:10.1126/sciadv.aax1950}, which proceeds as follows until the solution forms a clique. From the subgraph defined by the solution $\boldsymbol{x}$, we select the vertex with the smallest degree. In cases of degeneracy, we select the vertex with the lowest weight, breaking any further ties randomly. We then set the corresponding variable $i$ to $x_i=0$ and repeat the process until $\boldsymbol{x}$ is a valid clique.

Starting from this valid clique $\boldsymbol{x}$, we attempt to increase its weighted size using the local search strategy of Ref.~\cite{doi:10.1126/sciadv.aax1950}. For a fixed number of iterations, the algorithm proceeds as follows. First, we generate a pool of external vertices connected to all vertices currently in the clique. If this pool is non-empty, we select a vertex $i$ with probability $\propto h_i$, add it to the clique, and restart the process. If the pool is empty, we instead identify external vertices connected to all but one clique member. From this set, we select a vertex $i$ with probability $\propto h_i$) and swap it with the unconnected clique member, before returning to the first step.

\bibliography{bibliography}

\begin{thebibliography}{80}%
\makeatletter
\providecommand \@ifxundefined [1]{%
 \@ifx{#1\undefined}
}%
\providecommand \@ifnum [1]{%
 \ifnum #1\expandafter \@firstoftwo
 \else \expandafter \@secondoftwo
 \fi
}%
\providecommand \@ifx [1]{%
 \ifx #1\expandafter \@firstoftwo
 \else \expandafter \@secondoftwo
 \fi
}%
\providecommand \natexlab [1]{#1}%
\providecommand \enquote  [1]{``#1''}%
\providecommand \bibnamefont  [1]{#1}%
\providecommand \bibfnamefont [1]{#1}%
\providecommand \citenamefont [1]{#1}%
\providecommand \href@noop [0]{\@secondoftwo}%
\providecommand \href [0]{\begingroup \@sanitize@url \@href}%
\providecommand \@href[1]{\@@startlink{#1}\@@href}%
\providecommand \@@href[1]{\endgroup#1\@@endlink}%
\providecommand \@sanitize@url [0]{\catcode `\\12\catcode `\$12\catcode `\&12\catcode `\#12\catcode `\^12\catcode `\_12\catcode `\%12\relax}%
\providecommand \@@startlink[1]{}%
\providecommand \@@endlink[0]{}%
\providecommand \url  [0]{\begingroup\@sanitize@url \@url }%
\providecommand \@url [1]{\endgroup\@href {#1}{\urlprefix }}%
\providecommand \urlprefix  [0]{URL }%
\providecommand \Eprint [0]{\href }%
\providecommand \doibase [0]{https://doi.org/}%
\providecommand \selectlanguage [0]{\@gobble}%
\providecommand \bibinfo  [0]{\@secondoftwo}%
\providecommand \bibfield  [0]{\@secondoftwo}%
\providecommand \translation [1]{[#1]}%
\providecommand \BibitemOpen [0]{}%
\providecommand \bibitemStop [0]{}%
\providecommand \bibitemNoStop [0]{.\EOS\space}%
\providecommand \EOS [0]{\spacefactor3000\relax}%
\providecommand \BibitemShut  [1]{\csname bibitem#1\endcsname}%
\let\auto@bib@innerbib\@empty
\bibitem [{\citenamefont {Steane}(1998)}]{AndrewSteane_1998}%
  \BibitemOpen
  \bibfield  {author} {\bibinfo {author} {\bibfnamefont {A.}~\bibnamefont {Steane}},\ }\bibfield  {title} {\bibinfo {title} {Quantum computing},\ }\href {https://doi.org/10.1088/0034-4885/61/2/002} {\bibfield  {journal} {\bibinfo  {journal} {Rep. Prog. Phys.}\ }\textbf {\bibinfo {volume} {61}},\ \bibinfo {pages} {117} (\bibinfo {year} {1998})}\BibitemShut {NoStop}%
\bibitem [{\citenamefont {Ladd}\ \emph {et~al.}(2010)\citenamefont {Ladd}, \citenamefont {Jelezko}, \citenamefont {Laflamme}, \citenamefont {Nakamura}, \citenamefont {Monroe},\ and\ \citenamefont {O'Brien}}]{Ladd2010}%
  \BibitemOpen
  \bibfield  {author} {\bibinfo {author} {\bibfnamefont {T.~D.}\ \bibnamefont {Ladd}}, \bibinfo {author} {\bibfnamefont {F.}~\bibnamefont {Jelezko}}, \bibinfo {author} {\bibfnamefont {R.}~\bibnamefont {Laflamme}}, \bibinfo {author} {\bibfnamefont {Y.}~\bibnamefont {Nakamura}}, \bibinfo {author} {\bibfnamefont {C.}~\bibnamefont {Monroe}},\ and\ \bibinfo {author} {\bibfnamefont {J.~L.}\ \bibnamefont {O'Brien}},\ }\bibfield  {title} {\bibinfo {title} {Quantum computers},\ }\href {https://doi.org/10.1038/nature08812} {\bibfield  {journal} {\bibinfo  {journal} {Nature}\ }\textbf {\bibinfo {volume} {464}},\ \bibinfo {pages} {45} (\bibinfo {year} {2010})}\BibitemShut {NoStop}%
\bibitem [{\citenamefont {Shor}(1997)}]{shor1997polynomial}%
  \BibitemOpen
  \bibfield  {author} {\bibinfo {author} {\bibfnamefont {P.~W.}\ \bibnamefont {Shor}},\ }\bibfield  {title} {\bibinfo {title} {{Polynomial-Time Algorithms for Prime Factorization and Discrete Logarithms on a Quantum Computer}},\ }\href {https://doi.org/10.1137/S0097539795293172} {\bibfield  {journal} {\bibinfo  {journal} {SIAM J. Comput.}\ }\textbf {\bibinfo {volume} {26}},\ \bibinfo {pages} {1484} (\bibinfo {year} {1997})}\BibitemShut {NoStop}%
\bibitem [{\citenamefont {Kitaev}(1995)}]{kitaev1995quantum}%
  \BibitemOpen
  \bibfield  {author} {\bibinfo {author} {\bibfnamefont {A.~Y.}\ \bibnamefont {Kitaev}},\ }\bibfield  {title} {\bibinfo {title} {{Quantum measurements and the Abelian Stabilizer Problem}},\ }\href {https://arxiv.org/abs/quant-ph/9511026} {\bibfield  {journal} {\bibinfo  {journal} {arXiv:quant-ph/9511026}\ } (\bibinfo {year} {1995})}\BibitemShut {NoStop}%
\bibitem [{\citenamefont {Cleve}\ \emph {et~al.}(1998)\citenamefont {Cleve}, \citenamefont {Ekert}, \citenamefont {Macchiavello},\ and\ \citenamefont {Mosca}}]{cleve1998quantum}%
  \BibitemOpen
  \bibfield  {author} {\bibinfo {author} {\bibfnamefont {R.}~\bibnamefont {Cleve}}, \bibinfo {author} {\bibfnamefont {A.}~\bibnamefont {Ekert}}, \bibinfo {author} {\bibfnamefont {C.}~\bibnamefont {Macchiavello}},\ and\ \bibinfo {author} {\bibfnamefont {M.}~\bibnamefont {Mosca}},\ }\bibfield  {title} {\bibinfo {title} {Quantum algorithms revisited},\ }\href {https://doi.org/10.1098/rspa.1998.0164} {\bibfield  {journal} {\bibinfo  {journal} {Proceedings of the Royal Society of London. Series A: Mathematical, Physical and Engineering Sciences}\ }\textbf {\bibinfo {volume} {454}},\ \bibinfo {pages} {339} (\bibinfo {year} {1998})}\BibitemShut {NoStop}%
\bibitem [{\citenamefont {Cerezo}\ \emph {et~al.}(2022)\citenamefont {Cerezo}, \citenamefont {Verdon}, \citenamefont {Huang}, \citenamefont {Cincio},\ and\ \citenamefont {Coles}}]{cerezo2022challenges}%
  \BibitemOpen
  \bibfield  {author} {\bibinfo {author} {\bibfnamefont {M.}~\bibnamefont {Cerezo}}, \bibinfo {author} {\bibfnamefont {G.}~\bibnamefont {Verdon}}, \bibinfo {author} {\bibfnamefont {H.-Y.}\ \bibnamefont {Huang}}, \bibinfo {author} {\bibfnamefont {L.}~\bibnamefont {Cincio}},\ and\ \bibinfo {author} {\bibfnamefont {P.~J.}\ \bibnamefont {Coles}},\ }\bibfield  {title} {\bibinfo {title} {Challenges and opportunities in quantum machine learning},\ }\href {https://doi.org/10.1038/s43588-022-00311-3} {\bibfield  {journal} {\bibinfo  {journal} {Nat. Comput. Sci.}\ }\textbf {\bibinfo {volume} {2}},\ \bibinfo {pages} {567} (\bibinfo {year} {2022})}\BibitemShut {NoStop}%
\bibitem [{\citenamefont {Farhi}\ \emph {et~al.}(2014)\citenamefont {Farhi}, \citenamefont {Goldstone},\ and\ \citenamefont {Gutmann}}]{Farhi2014}%
  \BibitemOpen
  \bibfield  {author} {\bibinfo {author} {\bibfnamefont {E.}~\bibnamefont {Farhi}}, \bibinfo {author} {\bibfnamefont {J.}~\bibnamefont {Goldstone}},\ and\ \bibinfo {author} {\bibfnamefont {S.}~\bibnamefont {Gutmann}},\ }\bibfield  {title} {\bibinfo {title} {{A} {Q}uantum {A}pproximate {O}ptimization {A}lgorithm},\ }\href {https://arxiv.org/abs/1411.4028} {\bibfield  {journal} {\bibinfo  {journal} {arXiv:1411.4028}\ } (\bibinfo {year} {2014})}\BibitemShut {NoStop}%
\bibitem [{\citenamefont {Blekos}\ \emph {et~al.}(2024)\citenamefont {Blekos}, \citenamefont {Brand}, \citenamefont {Ceschini}, \citenamefont {Chou}, \citenamefont {Li}, \citenamefont {Pandya},\ and\ \citenamefont {Summer}}]{BLEKOS20241}%
  \BibitemOpen
  \bibfield  {author} {\bibinfo {author} {\bibfnamefont {K.}~\bibnamefont {Blekos}}, \bibinfo {author} {\bibfnamefont {D.}~\bibnamefont {Brand}}, \bibinfo {author} {\bibfnamefont {A.}~\bibnamefont {Ceschini}}, \bibinfo {author} {\bibfnamefont {C.-H.}\ \bibnamefont {Chou}}, \bibinfo {author} {\bibfnamefont {R.-H.}\ \bibnamefont {Li}}, \bibinfo {author} {\bibfnamefont {K.}~\bibnamefont {Pandya}},\ and\ \bibinfo {author} {\bibfnamefont {A.}~\bibnamefont {Summer}},\ }\bibfield  {title} {\bibinfo {title} {A review on quantum approximate optimization algorithm and its variants},\ }\href {https://doi.org/https://doi.org/10.1016/j.physrep.2024.03.002} {\bibfield  {journal} {\bibinfo  {journal} {Phys. Rep.}\ }\textbf {\bibinfo {volume} {1068}},\ \bibinfo {pages} {1} (\bibinfo {year} {2024})}\BibitemShut {NoStop}%
\bibitem [{\citenamefont {Ebadi}\ \emph {et~al.}(2022)\citenamefont {Ebadi}, \citenamefont {Keesling}, \citenamefont {Cain}, \citenamefont {Wang}, \citenamefont {Levine}, \citenamefont {Bluvstein}, \citenamefont {Semeghini}, \citenamefont {Omran}, \citenamefont {Liu}, \citenamefont {Samajdar}, \citenamefont {Luo}, \citenamefont {Nash}, \citenamefont {Gao}, \citenamefont {Barak}, \citenamefont {Farhi}, \citenamefont {Sachdev}, \citenamefont {Gemelke}, \citenamefont {Zhou}, \citenamefont {Choi}, \citenamefont {Pichler}, \citenamefont {Wang}, \citenamefont {Greiner}, \citenamefont {Vuletić},\ and\ \citenamefont {Lukin}}]{Ebadi2022}%
  \BibitemOpen
  \bibfield  {author} {\bibinfo {author} {\bibfnamefont {S.}~\bibnamefont {Ebadi}}, \bibinfo {author} {\bibfnamefont {A.}~\bibnamefont {Keesling}}, \bibinfo {author} {\bibfnamefont {M.}~\bibnamefont {Cain}}, \bibinfo {author} {\bibfnamefont {T.~T.}\ \bibnamefont {Wang}}, \bibinfo {author} {\bibfnamefont {H.}~\bibnamefont {Levine}}, \bibinfo {author} {\bibfnamefont {D.}~\bibnamefont {Bluvstein}}, \bibinfo {author} {\bibfnamefont {G.}~\bibnamefont {Semeghini}}, \bibinfo {author} {\bibfnamefont {A.}~\bibnamefont {Omran}}, \bibinfo {author} {\bibfnamefont {J.-G.}\ \bibnamefont {Liu}}, \bibinfo {author} {\bibfnamefont {R.}~\bibnamefont {Samajdar}}, \bibinfo {author} {\bibfnamefont {X.-Z.}\ \bibnamefont {Luo}}, \bibinfo {author} {\bibfnamefont {B.}~\bibnamefont {Nash}}, \bibinfo {author} {\bibfnamefont {X.}~\bibnamefont {Gao}}, \bibinfo {author} {\bibfnamefont {B.}~\bibnamefont {Barak}}, \bibinfo {author} {\bibfnamefont {E.}~\bibnamefont {Farhi}}, \bibinfo {author} {\bibfnamefont {S.}~\bibnamefont {Sachdev}},
  \bibinfo {author} {\bibfnamefont {N.}~\bibnamefont {Gemelke}}, \bibinfo {author} {\bibfnamefont {L.}~\bibnamefont {Zhou}}, \bibinfo {author} {\bibfnamefont {S.}~\bibnamefont {Choi}}, \bibinfo {author} {\bibfnamefont {H.}~\bibnamefont {Pichler}}, \bibinfo {author} {\bibfnamefont {S.-T.}\ \bibnamefont {Wang}}, \bibinfo {author} {\bibfnamefont {M.}~\bibnamefont {Greiner}}, \bibinfo {author} {\bibfnamefont {V.}~\bibnamefont {Vuletić}},\ and\ \bibinfo {author} {\bibfnamefont {M.~D.}\ \bibnamefont {Lukin}},\ }\bibfield  {title} {\bibinfo {title} {Quantum optimization of maximum independent set using rydberg atom arrays},\ }\href {https://doi.org/10.1126/science.abo6587} {\bibfield  {journal} {\bibinfo  {journal} {Science}\ }\textbf {\bibinfo {volume} {376}},\ \bibinfo {pages} {1209} (\bibinfo {year} {2022})}\BibitemShut {NoStop}%
\bibitem [{\citenamefont {Shaydulin}\ \emph {et~al.}(2024)\citenamefont {Shaydulin}, \citenamefont {Li}, \citenamefont {Chakrabarti}, \citenamefont {DeCross}, \citenamefont {Herman}, \citenamefont {Kumar}, \citenamefont {Larson}, \citenamefont {Lykov}, \citenamefont {Minssen}, \citenamefont {Sun}, \citenamefont {Alexeev}, \citenamefont {Dreiling}, \citenamefont {Gaebler}, \citenamefont {Gatterman}, \citenamefont {Gerber}, \citenamefont {Gilmore}, \citenamefont {Gresh}, \citenamefont {Hewitt}, \citenamefont {Horst}, \citenamefont {Hu}, \citenamefont {Johansen}, \citenamefont {Matheny}, \citenamefont {Mengle}, \citenamefont {Mills}, \citenamefont {Moses}, \citenamefont {Neyenhuis}, \citenamefont {Siegfried}, \citenamefont {Yalovetzky},\ and\ \citenamefont {Pistoia}}]{shaydulin2024evidence}%
  \BibitemOpen
  \bibfield  {author} {\bibinfo {author} {\bibfnamefont {R.}~\bibnamefont {Shaydulin}}, \bibinfo {author} {\bibfnamefont {C.}~\bibnamefont {Li}}, \bibinfo {author} {\bibfnamefont {S.}~\bibnamefont {Chakrabarti}}, \bibinfo {author} {\bibfnamefont {M.}~\bibnamefont {DeCross}}, \bibinfo {author} {\bibfnamefont {D.}~\bibnamefont {Herman}}, \bibinfo {author} {\bibfnamefont {N.}~\bibnamefont {Kumar}}, \bibinfo {author} {\bibfnamefont {J.}~\bibnamefont {Larson}}, \bibinfo {author} {\bibfnamefont {D.}~\bibnamefont {Lykov}}, \bibinfo {author} {\bibfnamefont {P.}~\bibnamefont {Minssen}}, \bibinfo {author} {\bibfnamefont {Y.}~\bibnamefont {Sun}}, \bibinfo {author} {\bibfnamefont {Y.}~\bibnamefont {Alexeev}}, \bibinfo {author} {\bibfnamefont {J.~M.}\ \bibnamefont {Dreiling}}, \bibinfo {author} {\bibfnamefont {J.~P.}\ \bibnamefont {Gaebler}}, \bibinfo {author} {\bibfnamefont {T.~M.}\ \bibnamefont {Gatterman}}, \bibinfo {author} {\bibfnamefont {J.~A.}\ \bibnamefont {Gerber}}, \bibinfo {author} {\bibfnamefont
  {K.}~\bibnamefont {Gilmore}}, \bibinfo {author} {\bibfnamefont {D.}~\bibnamefont {Gresh}}, \bibinfo {author} {\bibfnamefont {N.}~\bibnamefont {Hewitt}}, \bibinfo {author} {\bibfnamefont {C.~V.}\ \bibnamefont {Horst}}, \bibinfo {author} {\bibfnamefont {S.}~\bibnamefont {Hu}}, \bibinfo {author} {\bibfnamefont {J.}~\bibnamefont {Johansen}}, \bibinfo {author} {\bibfnamefont {M.}~\bibnamefont {Matheny}}, \bibinfo {author} {\bibfnamefont {T.}~\bibnamefont {Mengle}}, \bibinfo {author} {\bibfnamefont {M.}~\bibnamefont {Mills}}, \bibinfo {author} {\bibfnamefont {S.~A.}\ \bibnamefont {Moses}}, \bibinfo {author} {\bibfnamefont {B.}~\bibnamefont {Neyenhuis}}, \bibinfo {author} {\bibfnamefont {P.}~\bibnamefont {Siegfried}}, \bibinfo {author} {\bibfnamefont {R.}~\bibnamefont {Yalovetzky}},\ and\ \bibinfo {author} {\bibfnamefont {M.}~\bibnamefont {Pistoia}},\ }\bibfield  {title} {\bibinfo {title} {Evidence of scaling advantage for the quantum approximate optimization algorithm on a classically intractable problem},\
  }\href {https://doi.org/10.1126/sciadv.adm6761} {\bibfield  {journal} {\bibinfo  {journal} {Sci. Adv.}\ }\textbf {\bibinfo {volume} {10}},\ \bibinfo {pages} {eadm6761} (\bibinfo {year} {2024})}\BibitemShut {NoStop}%
\bibitem [{\citenamefont {Cadavid}\ \emph {et~al.}(2025)\citenamefont {Cadavid}, \citenamefont {Chandarana}, \citenamefont {Romero}, \citenamefont {Trautmann}, \citenamefont {Solano}, \citenamefont {Patti},\ and\ \citenamefont {Hegade}}]{Cadavid2025}%
  \BibitemOpen
  \bibfield  {author} {\bibinfo {author} {\bibfnamefont {A.~G.}\ \bibnamefont {Cadavid}}, \bibinfo {author} {\bibfnamefont {P.}~\bibnamefont {Chandarana}}, \bibinfo {author} {\bibfnamefont {S.~V.}\ \bibnamefont {Romero}}, \bibinfo {author} {\bibfnamefont {J.}~\bibnamefont {Trautmann}}, \bibinfo {author} {\bibfnamefont {E.}~\bibnamefont {Solano}}, \bibinfo {author} {\bibfnamefont {T.~L.}\ \bibnamefont {Patti}},\ and\ \bibinfo {author} {\bibfnamefont {N.~N.}\ \bibnamefont {Hegade}},\ }\bibfield  {title} {\bibinfo {title} {{Scaling advantage with quantum-enhanced memetic tabu search for LABS}},\ }\href {https://arxiv.org/abs/2511.04553} {\bibfield  {journal} {\bibinfo  {journal} {arXiv:2511.04553}\ } (\bibinfo {year} {2025})}\BibitemShut {NoStop}%
\bibitem [{\citenamefont {Albash}\ and\ \citenamefont {Lidar}(2018)}]{PhysRevX.8.031016}%
  \BibitemOpen
  \bibfield  {author} {\bibinfo {author} {\bibfnamefont {T.}~\bibnamefont {Albash}}\ and\ \bibinfo {author} {\bibfnamefont {D.~A.}\ \bibnamefont {Lidar}},\ }\bibfield  {title} {\bibinfo {title} {Demonstration of a scaling advantage for a quantum annealer over simulated annealing},\ }\href {https://doi.org/10.1103/PhysRevX.8.031016} {\bibfield  {journal} {\bibinfo  {journal} {Phys. Rev. X}\ }\textbf {\bibinfo {volume} {8}},\ \bibinfo {pages} {031016} (\bibinfo {year} {2018})}\BibitemShut {NoStop}%
\bibitem [{\citenamefont {Munoz-Bauza}\ and\ \citenamefont {Lidar}(2025)}]{Bauza2024}%
  \BibitemOpen
  \bibfield  {author} {\bibinfo {author} {\bibfnamefont {H.}~\bibnamefont {Munoz-Bauza}}\ and\ \bibinfo {author} {\bibfnamefont {D.}~\bibnamefont {Lidar}},\ }\bibfield  {title} {\bibinfo {title} {Scaling advantage in approximate optimization with quantum annealing},\ }\href {https://doi.org/10.1103/PhysRevLett.134.160601} {\bibfield  {journal} {\bibinfo  {journal} {Phys. Rev. Lett.}\ }\textbf {\bibinfo {volume} {134}},\ \bibinfo {pages} {160601} (\bibinfo {year} {2025})}\BibitemShut {NoStop}%
\bibitem [{\citenamefont {Chandarana}\ \emph {et~al.}(2025{\natexlab{a}})\citenamefont {Chandarana}, \citenamefont {Cadavid}, \citenamefont {Romero}, \citenamefont {Anton~Simen},\ and\ \citenamefont {Hegade}}]{Chandarana2025}%
  \BibitemOpen
  \bibfield  {author} {\bibinfo {author} {\bibfnamefont {P.}~\bibnamefont {Chandarana}}, \bibinfo {author} {\bibfnamefont {A.~G.}\ \bibnamefont {Cadavid}}, \bibinfo {author} {\bibfnamefont {S.~V.}\ \bibnamefont {Romero}}, \bibinfo {author} {\bibfnamefont {E.~S.}\ \bibnamefont {Anton~Simen}},\ and\ \bibinfo {author} {\bibfnamefont {N.~N.}\ \bibnamefont {Hegade}},\ }\bibfield  {title} {\bibinfo {title} {{Runtime Quantum Advantage with Digital Quantum Optimization }},\ }\href {https://arxiv.org/abs/2505.08663} {\bibfield  {journal} {\bibinfo  {journal} {arXiv:2505.08663}\ } (\bibinfo {year} {2025}{\natexlab{a}})}\BibitemShut {NoStop}%
\bibitem [{\citenamefont {Chandarana}\ \emph {et~al.}(2025{\natexlab{b}})\citenamefont {Chandarana}, \citenamefont {Romero}, \citenamefont {Cadavid}, \citenamefont {Simen}, \citenamefont {Solano},\ and\ \citenamefont {Hegade}}]{Chandarana2025b}%
  \BibitemOpen
  \bibfield  {author} {\bibinfo {author} {\bibfnamefont {P.}~\bibnamefont {Chandarana}}, \bibinfo {author} {\bibfnamefont {S.~V.}\ \bibnamefont {Romero}}, \bibinfo {author} {\bibfnamefont {A.~G.}\ \bibnamefont {Cadavid}}, \bibinfo {author} {\bibfnamefont {A.}~\bibnamefont {Simen}}, \bibinfo {author} {\bibfnamefont {E.}~\bibnamefont {Solano}},\ and\ \bibinfo {author} {\bibfnamefont {N.~N.}\ \bibnamefont {Hegade}},\ }\bibfield  {title} {\bibinfo {title} {{Hybrid Sequential Quantum Computing}},\ }\href {https://arxiv.org/abs/2510.05851} {\bibfield  {journal} {\bibinfo  {journal} {arXiv:2510.05851}\ } (\bibinfo {year} {2025}{\natexlab{b}})}\BibitemShut {NoStop}%
\bibitem [{\citenamefont {Farr\'e'}\ \emph {et~al.}(2025)\citenamefont {Farr\'e'}, \citenamefont {Ordog}, \citenamefont {Chern},\ and\ \citenamefont {McGeoch}}]{Farre2025}%
  \BibitemOpen
  \bibfield  {author} {\bibinfo {author} {\bibfnamefont {P.}~\bibnamefont {Farr\'e'}}, \bibinfo {author} {\bibfnamefont {E.}~\bibnamefont {Ordog}}, \bibinfo {author} {\bibfnamefont {K.}~\bibnamefont {Chern}},\ and\ \bibinfo {author} {\bibfnamefont {C.~C.}\ \bibnamefont {McGeoch}},\ }\bibfield  {title} {\bibinfo {title} {{Comparing Quantum Annealing and BF-DCQO}},\ }\href {https://arxiv.org/abs/2509.14358} {\bibfield  {journal} {\bibinfo  {journal} {arXiv:2509.14358}\ } (\bibinfo {year} {2025})}\BibitemShut {NoStop}%
\bibitem [{\citenamefont {Tuziemski}\ \emph {et~al.}(2025)\citenamefont {Tuziemski}, \citenamefont {Paw\l{}owski}, \citenamefont {Tarasiuk}, \citenamefont {Pawela},\ and\ \citenamefont {Gardas}}]{Tuziemski2025}%
  \BibitemOpen
  \bibfield  {author} {\bibinfo {author} {\bibfnamefont {J.}~\bibnamefont {Tuziemski}}, \bibinfo {author} {\bibfnamefont {J.}~\bibnamefont {Paw\l{}owski}}, \bibinfo {author} {\bibfnamefont {P.}~\bibnamefont {Tarasiuk}}, \bibinfo {author} {\bibfnamefont {L.}~\bibnamefont {Pawela}},\ and\ \bibinfo {author} {\bibfnamefont {B.}~\bibnamefont {Gardas}},\ }\bibfield  {title} {\bibinfo {title} {{Recent quantum runtime (dis)advantages}},\ }\href {https://arxiv.org/abs/2510.06337} {\bibfield  {journal} {\bibinfo  {journal} {arXiv:2510.06337}\ } (\bibinfo {year} {2025})}\BibitemShut {NoStop}%
\bibitem [{\citenamefont {Patti}\ \emph {et~al.}(2022{\natexlab{a}})\citenamefont {Patti}, \citenamefont {Kossaifi}, \citenamefont {Anandkumar},\ and\ \citenamefont {Yelin}}]{patti2022variational}%
  \BibitemOpen
  \bibfield  {author} {\bibinfo {author} {\bibfnamefont {T.~L.}\ \bibnamefont {Patti}}, \bibinfo {author} {\bibfnamefont {J.}~\bibnamefont {Kossaifi}}, \bibinfo {author} {\bibfnamefont {A.}~\bibnamefont {Anandkumar}},\ and\ \bibinfo {author} {\bibfnamefont {S.~F.}\ \bibnamefont {Yelin}},\ }\bibfield  {title} {\bibinfo {title} {Variational quantum optimization with multibasis encodings},\ }\href {https://doi.org/10.1103/PhysRevResearch.4.033142} {\bibfield  {journal} {\bibinfo  {journal} {Phys. Rev. Res.}\ }\textbf {\bibinfo {volume} {4}},\ \bibinfo {pages} {033142} (\bibinfo {year} {2022}{\natexlab{a}})}\BibitemShut {NoStop}%
\bibitem [{\citenamefont {Tene-Cohen}\ \emph {et~al.}(2023)\citenamefont {Tene-Cohen}, \citenamefont {Kelman}, \citenamefont {Lev},\ and\ \citenamefont {Makmal}}]{Tenecohen2023}%
  \BibitemOpen
  \bibfield  {author} {\bibinfo {author} {\bibfnamefont {Y.}~\bibnamefont {Tene-Cohen}}, \bibinfo {author} {\bibfnamefont {T.}~\bibnamefont {Kelman}}, \bibinfo {author} {\bibfnamefont {O.}~\bibnamefont {Lev}},\ and\ \bibinfo {author} {\bibfnamefont {A.}~\bibnamefont {Makmal}},\ }\bibfield  {title} {\bibinfo {title} {{A Variational Qubit-Efficient MaxCut Heuristic Algorithm}},\ }\href {https://arxiv.org/abs/2308.10383} {\bibfield  {journal} {\bibinfo  {journal} {arXiv:2308.10383}\ } (\bibinfo {year} {2023})}\BibitemShut {NoStop}%
\bibitem [{\citenamefont {Ran\ifmmode \check{c}\else \v{c}\fi{}i\ifmmode~\acute{c}\else \'{c}\fi{}}(2023)}]{PhysRevResearch.5.L012021}%
  \BibitemOpen
  \bibfield  {author} {\bibinfo {author} {\bibfnamefont {M.~J.}\ \bibnamefont {Ran\ifmmode \check{c}\else \v{c}\fi{}i\ifmmode~\acute{c}\else \'{c}\fi{}}},\ }\bibfield  {title} {\bibinfo {title} {Noisy intermediate-scale quantum computing algorithm for solving an $n$-vertex maxcut problem with log($n$) qubits},\ }\href {https://doi.org/10.1103/PhysRevResearch.5.L012021} {\bibfield  {journal} {\bibinfo  {journal} {Phys. Rev. Res.}\ }\textbf {\bibinfo {volume} {5}},\ \bibinfo {pages} {L012021} (\bibinfo {year} {2023})}\BibitemShut {NoStop}%
\bibitem [{\citenamefont {Chatterjee}\ \emph {et~al.}(2024)\citenamefont {Chatterjee}, \citenamefont {Bourreau},\ and\ \citenamefont {Ran\ifmmode \check{c}\else \v{c}\fi{}i\ifmmode~\acute{c}\else \'{c}\fi{}}}]{PhysRevA.109.052441}%
  \BibitemOpen
  \bibfield  {author} {\bibinfo {author} {\bibfnamefont {Y.}~\bibnamefont {Chatterjee}}, \bibinfo {author} {\bibfnamefont {E.}~\bibnamefont {Bourreau}},\ and\ \bibinfo {author} {\bibfnamefont {M.~J.}\ \bibnamefont {Ran\ifmmode \check{c}\else \v{c}\fi{}i\ifmmode~\acute{c}\else \'{c}\fi{}}},\ }\bibfield  {title} {\bibinfo {title} {Solving various np-hard problems using exponentially fewer qubits on a quantum computer},\ }\href {https://doi.org/10.1103/PhysRevA.109.052441} {\bibfield  {journal} {\bibinfo  {journal} {Phys. Rev. A}\ }\textbf {\bibinfo {volume} {109}},\ \bibinfo {pages} {052441} (\bibinfo {year} {2024})}\BibitemShut {NoStop}%
\bibitem [{\citenamefont {Patti}\ \emph {et~al.}(2022{\natexlab{b}})\citenamefont {Patti}, \citenamefont {Kossaifi}, \citenamefont {Anandkumar},\ and\ \citenamefont {Yelin}}]{PhysRevResearch.4.033142}%
  \BibitemOpen
  \bibfield  {author} {\bibinfo {author} {\bibfnamefont {T.~L.}\ \bibnamefont {Patti}}, \bibinfo {author} {\bibfnamefont {J.}~\bibnamefont {Kossaifi}}, \bibinfo {author} {\bibfnamefont {A.}~\bibnamefont {Anandkumar}},\ and\ \bibinfo {author} {\bibfnamefont {S.~F.}\ \bibnamefont {Yelin}},\ }\bibfield  {title} {\bibinfo {title} {Variational quantum optimization with multibasis encodings},\ }\href {https://doi.org/10.1103/PhysRevResearch.4.033142} {\bibfield  {journal} {\bibinfo  {journal} {Phys. Rev. Res.}\ }\textbf {\bibinfo {volume} {4}},\ \bibinfo {pages} {033142} (\bibinfo {year} {2022}{\natexlab{b}})}\BibitemShut {NoStop}%
\bibitem [{\citenamefont {Sciorilli}\ \emph {et~al.}(2025)\citenamefont {Sciorilli}, \citenamefont {Borges}, \citenamefont {Patti}, \citenamefont {Garc{\'\i}a-Mart{\'\i}n}, \citenamefont {Camilo}, \citenamefont {Anandkumar},\ and\ \citenamefont {Aolita}}]{Sciorilli2024}%
  \BibitemOpen
  \bibfield  {author} {\bibinfo {author} {\bibfnamefont {M.}~\bibnamefont {Sciorilli}}, \bibinfo {author} {\bibfnamefont {L.}~\bibnamefont {Borges}}, \bibinfo {author} {\bibfnamefont {T.~L.}\ \bibnamefont {Patti}}, \bibinfo {author} {\bibfnamefont {D.}~\bibnamefont {Garc{\'\i}a-Mart{\'\i}n}}, \bibinfo {author} {\bibfnamefont {G.}~\bibnamefont {Camilo}}, \bibinfo {author} {\bibfnamefont {A.}~\bibnamefont {Anandkumar}},\ and\ \bibinfo {author} {\bibfnamefont {L.}~\bibnamefont {Aolita}},\ }\bibfield  {title} {\bibinfo {title} {Towards large-scale quantum optimization solvers with few qubits},\ }\href {https://doi.org/10.1038/s41467-024-55346-z} {\bibfield  {journal} {\bibinfo  {journal} {Nat. Comm.}\ }\textbf {\bibinfo {volume} {16}},\ \bibinfo {pages} {476} (\bibinfo {year} {2025})}\BibitemShut {NoStop}%
\bibitem [{\citenamefont {Sundar}\ and\ \citenamefont {Dupont}(2024)}]{Sundar2024}%
  \BibitemOpen
  \bibfield  {author} {\bibinfo {author} {\bibfnamefont {B.}~\bibnamefont {Sundar}}\ and\ \bibinfo {author} {\bibfnamefont {M.}~\bibnamefont {Dupont}},\ }\bibfield  {title} {\bibinfo {title} {Qubit-efficient quantum combinatorial optimization solver},\ }\href {https://arxiv.org/abs/2407.15539} {\bibfield  {journal} {\bibinfo  {journal} {arXiv:2407.15539}\ } (\bibinfo {year} {2024})}\BibitemShut {NoStop}%
\bibitem [{\citenamefont {Bechtold}\ \emph {et~al.}(2023)\citenamefont {Bechtold}, \citenamefont {Barzen}, \citenamefont {Leymann}, \citenamefont {Mandl}, \citenamefont {Obst}, \citenamefont {Truger},\ and\ \citenamefont {Weder}}]{Bechtold_2023}%
  \BibitemOpen
  \bibfield  {author} {\bibinfo {author} {\bibfnamefont {M.}~\bibnamefont {Bechtold}}, \bibinfo {author} {\bibfnamefont {J.}~\bibnamefont {Barzen}}, \bibinfo {author} {\bibfnamefont {F.}~\bibnamefont {Leymann}}, \bibinfo {author} {\bibfnamefont {A.}~\bibnamefont {Mandl}}, \bibinfo {author} {\bibfnamefont {J.}~\bibnamefont {Obst}}, \bibinfo {author} {\bibfnamefont {F.}~\bibnamefont {Truger}},\ and\ \bibinfo {author} {\bibfnamefont {B.}~\bibnamefont {Weder}},\ }\bibfield  {title} {\bibinfo {title} {Investigating the effect of circuit cutting in qaoa for the maxcut problem on nisq devices},\ }\href {https://doi.org/10.1088/2058-9565/acf59c} {\bibfield  {journal} {\bibinfo  {journal} {Quantum Sci. Technol.}\ }\textbf {\bibinfo {volume} {8}},\ \bibinfo {pages} {045022} (\bibinfo {year} {2023})}\BibitemShut {NoStop}%
\bibitem [{\citenamefont {Dupont}\ \emph {et~al.}(2025{\natexlab{a}})\citenamefont {Dupont}, \citenamefont {Sundar}, \citenamefont {Evert}, \citenamefont {Neira}, \citenamefont {Peng}, \citenamefont {Jeffrey},\ and\ \citenamefont {Hodson}}]{PhysRevApplied.23.014045}%
  \BibitemOpen
  \bibfield  {author} {\bibinfo {author} {\bibfnamefont {M.}~\bibnamefont {Dupont}}, \bibinfo {author} {\bibfnamefont {B.}~\bibnamefont {Sundar}}, \bibinfo {author} {\bibfnamefont {B.}~\bibnamefont {Evert}}, \bibinfo {author} {\bibfnamefont {D.~E.~B.}\ \bibnamefont {Neira}}, \bibinfo {author} {\bibfnamefont {Z.}~\bibnamefont {Peng}}, \bibinfo {author} {\bibfnamefont {S.}~\bibnamefont {Jeffrey}},\ and\ \bibinfo {author} {\bibfnamefont {M.~J.}\ \bibnamefont {Hodson}},\ }\bibfield  {title} {\bibinfo {title} {Benchmarking quantum optimization for the maximum-cut problem on a superconducting quantum computer},\ }\href {https://doi.org/10.1103/PhysRevApplied.23.014045} {\bibfield  {journal} {\bibinfo  {journal} {Phys. Rev. Appl.}\ }\textbf {\bibinfo {volume} {23}},\ \bibinfo {pages} {014045} (\bibinfo {year} {2025}{\natexlab{a}})}\BibitemShut {NoStop}%
\bibitem [{\citenamefont {Dupont}\ \emph {et~al.}(2025{\natexlab{b}})\citenamefont {Dupont}, \citenamefont {Oberoi},\ and\ \citenamefont {Sundar}}]{9prw-684p}%
  \BibitemOpen
  \bibfield  {author} {\bibinfo {author} {\bibfnamefont {M.}~\bibnamefont {Dupont}}, \bibinfo {author} {\bibfnamefont {T.}~\bibnamefont {Oberoi}},\ and\ \bibinfo {author} {\bibfnamefont {B.}~\bibnamefont {Sundar}},\ }\bibfield  {title} {\bibinfo {title} {Optimization via quantum preconditioning},\ }\href {https://doi.org/10.1103/9prw-684p} {\bibfield  {journal} {\bibinfo  {journal} {Phys. Rev. Appl.}\ }\textbf {\bibinfo {volume} {24}},\ \bibinfo {pages} {044013} (\bibinfo {year} {2025}{\natexlab{b}})}\BibitemShut {NoStop}%
\bibitem [{\citenamefont {Liu}\ and\ \citenamefont {Goan}(2022)}]{Liu2022}%
  \BibitemOpen
  \bibfield  {author} {\bibinfo {author} {\bibfnamefont {C.-Y.}\ \bibnamefont {Liu}}\ and\ \bibinfo {author} {\bibfnamefont {H.-S.}\ \bibnamefont {Goan}},\ }\bibfield  {title} {\bibinfo {title} {{H}ybrid {G}ate-{B}ased and {A}nnealing {Q}uantum {C}omputing for {L}arge-{S}ize {I}sing {P}roblems},\ }\href {https://arxiv.org/abs/2208.03283} {\bibfield  {journal} {\bibinfo  {journal} {arXiv:2208.03283}\ } (\bibinfo {year} {2022})}\BibitemShut {NoStop}%
\bibitem [{\citenamefont {Tomesh}\ \emph {et~al.}(2023)\citenamefont {Tomesh}, \citenamefont {Saleem}, \citenamefont {Perlin}, \citenamefont {Gokhale}, \citenamefont {Suchara},\ and\ \citenamefont {Martonosi}}]{tomesh2023divide}%
  \BibitemOpen
  \bibfield  {author} {\bibinfo {author} {\bibfnamefont {T.}~\bibnamefont {Tomesh}}, \bibinfo {author} {\bibfnamefont {Z.~H.}\ \bibnamefont {Saleem}}, \bibinfo {author} {\bibfnamefont {M.~A.}\ \bibnamefont {Perlin}}, \bibinfo {author} {\bibfnamefont {P.}~\bibnamefont {Gokhale}}, \bibinfo {author} {\bibfnamefont {M.}~\bibnamefont {Suchara}},\ and\ \bibinfo {author} {\bibfnamefont {M.}~\bibnamefont {Martonosi}},\ }\bibfield  {title} {\bibinfo {title} {{ Divide and Conquer for Combinatorial Optimization and Distributed Quantum Computation }},\ }in\ \href {https://doi.org/10.1109/QCE57702.2023.00009} {\emph {\bibinfo {booktitle} {2023 IEEE International Conference on Quantum Computing and Engineering (QCE)}}}\ (\bibinfo  {publisher} {IEEE Computer Society},\ \bibinfo {address} {Los Alamitos, CA, USA},\ \bibinfo {year} {2023})\ pp.\ \bibinfo {pages} {1--12}\BibitemShut {NoStop}%
\bibitem [{\citenamefont {Zhou}\ \emph {et~al.}(2023)\citenamefont {Zhou}, \citenamefont {Du}, \citenamefont {Tian},\ and\ \citenamefont {Tao}}]{zhou2023qaoa}%
  \BibitemOpen
  \bibfield  {author} {\bibinfo {author} {\bibfnamefont {Z.}~\bibnamefont {Zhou}}, \bibinfo {author} {\bibfnamefont {Y.}~\bibnamefont {Du}}, \bibinfo {author} {\bibfnamefont {X.}~\bibnamefont {Tian}},\ and\ \bibinfo {author} {\bibfnamefont {D.}~\bibnamefont {Tao}},\ }\bibfield  {title} {\bibinfo {title} {Qaoa-in-qaoa: Solving large-scale maxcut problems on small quantum machines},\ }\href {https://doi.org/10.1103/PhysRevApplied.19.024027} {\bibfield  {journal} {\bibinfo  {journal} {Phys. Rev. Appl.}\ }\textbf {\bibinfo {volume} {19}},\ \bibinfo {pages} {024027} (\bibinfo {year} {2023})}\BibitemShut {NoStop}%
\bibitem [{\citenamefont {{\v{C}}epait{\.e}}\ \emph {et~al.}(2025)\citenamefont {{\v{C}}epait{\.e}}, \citenamefont {Vaishnav}, \citenamefont {Zhou},\ and\ \citenamefont {Montanaro}}]{vcepaite2025quantum}%
  \BibitemOpen
  \bibfield  {author} {\bibinfo {author} {\bibfnamefont {I.}~\bibnamefont {{\v{C}}epait{\.e}}}, \bibinfo {author} {\bibfnamefont {N.}~\bibnamefont {Vaishnav}}, \bibinfo {author} {\bibfnamefont {L.}~\bibnamefont {Zhou}},\ and\ \bibinfo {author} {\bibfnamefont {A.}~\bibnamefont {Montanaro}},\ }\bibfield  {title} {\bibinfo {title} {{Quantum-Enhanced Optimization by Warm Starts}},\ }\href {https://arxiv.org/abs/2508.16309} {\bibfield  {journal} {\bibinfo  {journal} {arXiv:2508.16309}\ } (\bibinfo {year} {2025})}\BibitemShut {NoStop}%
\bibitem [{\citenamefont {Huang}\ \emph {et~al.}(2026)\citenamefont {Huang}, \citenamefont {Jin}, \citenamefont {Zhang}, \citenamefont {Zhao}, \citenamefont {Qi},\ and\ \citenamefont {Shao}}]{Huang2026}%
  \BibitemOpen
  \bibfield  {author} {\bibinfo {author} {\bibfnamefont {Y.}~\bibnamefont {Huang}}, \bibinfo {author} {\bibfnamefont {S.}~\bibnamefont {Jin}}, \bibinfo {author} {\bibfnamefont {Y.}~\bibnamefont {Zhang}}, \bibinfo {author} {\bibfnamefont {Q.}~\bibnamefont {Zhao}}, \bibinfo {author} {\bibfnamefont {J.}~\bibnamefont {Qi}},\ and\ \bibinfo {author} {\bibfnamefont {Q.}~\bibnamefont {Shao}},\ }\bibfield  {title} {\bibinfo {title} {{Tensor Network Assisted Distributed Variational Quantum Algorithm for Large Scale Combinatorial Optimization Problem}},\ }\href {https://arxiv.org/abs/2601.13956} {\bibfield  {journal} {\bibinfo  {journal} {arXiv:2601.13956}\ } (\bibinfo {year} {2026})}\BibitemShut {NoStop}%
\bibitem [{\citenamefont {Ponce}\ \emph {et~al.}(2025)\citenamefont {Ponce}, \citenamefont {Herrman}, \citenamefont {Lotshaw}, \citenamefont {Powers}, \citenamefont {Siopsis}, \citenamefont {Humble},\ and\ \citenamefont {Ostrowski}}]{Ponce2023}%
  \BibitemOpen
  \bibfield  {author} {\bibinfo {author} {\bibfnamefont {M.}~\bibnamefont {Ponce}}, \bibinfo {author} {\bibfnamefont {R.}~\bibnamefont {Herrman}}, \bibinfo {author} {\bibfnamefont {P.~C.}\ \bibnamefont {Lotshaw}}, \bibinfo {author} {\bibfnamefont {S.}~\bibnamefont {Powers}}, \bibinfo {author} {\bibfnamefont {G.}~\bibnamefont {Siopsis}}, \bibinfo {author} {\bibfnamefont {T.}~\bibnamefont {Humble}},\ and\ \bibinfo {author} {\bibfnamefont {J.}~\bibnamefont {Ostrowski}},\ }\bibfield  {title} {\bibinfo {title} {Graph decomposition techniques for solving combinatorial optimization problems with variational quantum algorithms},\ }\href {https://doi.org/10.1007/s11128-025-04675-z} {\bibfield  {journal} {\bibinfo  {journal} {Quantum Inf. Process.}\ }\textbf {\bibinfo {volume} {24}},\ \bibinfo {pages} {60} (\bibinfo {year} {2025})}\BibitemShut {NoStop}%
\bibitem [{\citenamefont {Moondra}\ \emph {et~al.}(2024)\citenamefont {Moondra}, \citenamefont {Lotshaw}, \citenamefont {Mohler},\ and\ \citenamefont {Gupta}}]{Moondra2024}%
  \BibitemOpen
  \bibfield  {author} {\bibinfo {author} {\bibfnamefont {J.}~\bibnamefont {Moondra}}, \bibinfo {author} {\bibfnamefont {P.~C.}\ \bibnamefont {Lotshaw}}, \bibinfo {author} {\bibfnamefont {G.}~\bibnamefont {Mohler}},\ and\ \bibinfo {author} {\bibfnamefont {S.}~\bibnamefont {Gupta}},\ }\bibfield  {title} {\bibinfo {title} {{Promise of Graph Sparsification and Decomposition for Noise Reduction in QAOA: Analysis for Trapped-Ion Compilations}},\ }\href {https://arxiv.org/abs/2406.14330} {\bibfield  {journal} {\bibinfo  {journal} {arXiv:2406.14330}\ } (\bibinfo {year} {2024})}\BibitemShut {NoStop}%
\bibitem [{\citenamefont {Acharya}\ \emph {et~al.}(2025)\citenamefont {Acharya}, \citenamefont {Yalovetzky}, \citenamefont {Minssen}, \citenamefont {Chakrabarti}, \citenamefont {Shaydulin}, \citenamefont {Raymond}, \citenamefont {Sun}, \citenamefont {Herman}, \citenamefont {Andrist}, \citenamefont {Salton}, \citenamefont {Schuetz}, \citenamefont {Katzgraber},\ and\ \citenamefont {Pistoia}}]{Acharya2024}%
  \BibitemOpen
  \bibfield  {author} {\bibinfo {author} {\bibfnamefont {A.}~\bibnamefont {Acharya}}, \bibinfo {author} {\bibfnamefont {R.}~\bibnamefont {Yalovetzky}}, \bibinfo {author} {\bibfnamefont {P.}~\bibnamefont {Minssen}}, \bibinfo {author} {\bibfnamefont {S.}~\bibnamefont {Chakrabarti}}, \bibinfo {author} {\bibfnamefont {R.}~\bibnamefont {Shaydulin}}, \bibinfo {author} {\bibfnamefont {R.}~\bibnamefont {Raymond}}, \bibinfo {author} {\bibfnamefont {Y.}~\bibnamefont {Sun}}, \bibinfo {author} {\bibfnamefont {D.}~\bibnamefont {Herman}}, \bibinfo {author} {\bibfnamefont {R.~S.}\ \bibnamefont {Andrist}}, \bibinfo {author} {\bibfnamefont {G.}~\bibnamefont {Salton}}, \bibinfo {author} {\bibfnamefont {M.~J.~A.}\ \bibnamefont {Schuetz}}, \bibinfo {author} {\bibfnamefont {H.~G.}\ \bibnamefont {Katzgraber}},\ and\ \bibinfo {author} {\bibfnamefont {M.}~\bibnamefont {Pistoia}},\ }\bibfield  {title} {\bibinfo {title} {Decomposition pipeline for large-scale portfolio optimization with applications to near-term quantum computing},\
  }\href {https://doi.org/10.1103/PhysRevResearch.7.023142} {\bibfield  {journal} {\bibinfo  {journal} {Phys. Rev. Res.}\ }\textbf {\bibinfo {volume} {7}},\ \bibinfo {pages} {023142} (\bibinfo {year} {2025})}\BibitemShut {NoStop}%
\bibitem [{\citenamefont {Angone}\ \emph {et~al.}(2023)\citenamefont {Angone}, \citenamefont {Liu}, \citenamefont {Shaydulin},\ and\ \citenamefont {Safro}}]{10363584}%
  \BibitemOpen
  \bibfield  {author} {\bibinfo {author} {\bibfnamefont {A.}~\bibnamefont {Angone}}, \bibinfo {author} {\bibfnamefont {X.}~\bibnamefont {Liu}}, \bibinfo {author} {\bibfnamefont {R.}~\bibnamefont {Shaydulin}},\ and\ \bibinfo {author} {\bibfnamefont {I.}~\bibnamefont {Safro}},\ }\bibfield  {title} {\bibinfo {title} {Hybrid quantum-classical multilevel approach for maximum cuts on graphs},\ }in\ \href {https://doi.org/10.1109/HPEC58863.2023.10363584} {\emph {\bibinfo {booktitle} {2023 IEEE High Performance Extreme Computing Conference (HPEC)}}}\ (\bibinfo {year} {2023})\ pp.\ \bibinfo {pages} {1--7}\BibitemShut {NoStop}%
\bibitem [{\citenamefont {Bach}\ \emph {et~al.}(2024)\citenamefont {Bach}, \citenamefont {Falla},\ and\ \citenamefont {Safro}}]{Bach2024}%
  \BibitemOpen
  \bibfield  {author} {\bibinfo {author} {\bibfnamefont {B.}~\bibnamefont {Bach}}, \bibinfo {author} {\bibfnamefont {J.}~\bibnamefont {Falla}},\ and\ \bibinfo {author} {\bibfnamefont {I.}~\bibnamefont {Safro}},\ }\bibfield  {title} {\bibinfo {title} {Mlqaoa: Graph learning accelerated hybrid quantum-classical multilevel qaoa},\ }in\ \href {https://doi.org/10.1109/QCE60285.2024.00072} {\emph {\bibinfo {booktitle} {2024 IEEE International Conference on Quantum Computing and Engineering (QCE)}}},\ Vol.~\bibinfo {volume} {01}\ (\bibinfo {year} {2024})\ pp.\ \bibinfo {pages} {1--12}\BibitemShut {NoStop}%
\bibitem [{\citenamefont {Maciejewski}\ \emph {et~al.}(2024)\citenamefont {Maciejewski}, \citenamefont {Bach}, \citenamefont {Dupont}, \citenamefont {Lott}, \citenamefont {Sundar}, \citenamefont {Neira}, \citenamefont {Safro},\ and\ \citenamefont {Venturelli}}]{Maciejewski2024}%
  \BibitemOpen
  \bibfield  {author} {\bibinfo {author} {\bibfnamefont {F.~B.}\ \bibnamefont {Maciejewski}}, \bibinfo {author} {\bibfnamefont {B.~G.}\ \bibnamefont {Bach}}, \bibinfo {author} {\bibfnamefont {M.}~\bibnamefont {Dupont}}, \bibinfo {author} {\bibfnamefont {P.~A.}\ \bibnamefont {Lott}}, \bibinfo {author} {\bibfnamefont {B.}~\bibnamefont {Sundar}}, \bibinfo {author} {\bibfnamefont {D.~E.~B.}\ \bibnamefont {Neira}}, \bibinfo {author} {\bibfnamefont {I.}~\bibnamefont {Safro}},\ and\ \bibinfo {author} {\bibfnamefont {D.}~\bibnamefont {Venturelli}},\ }\bibfield  {title} {\bibinfo {title} {{A Multilevel Approach for Solving Large-Scale QUBO Problems with Noisy Hybrid Quantum Approximate Optimization}},\ }in\ \href {https://doi.org/10.1109/HPEC62836.2024.10938438} {\emph {\bibinfo {booktitle} {{2024 IEEE High Performance Extreme Computing Conference (HPEC)}}}}\ (\bibinfo {year} {2024})\ pp.\ \bibinfo {pages} {1--10}\BibitemShut {NoStop}%
\bibitem [{\citenamefont {Lucas}(2014)}]{lucas2014}%
  \BibitemOpen
  \bibfield  {author} {\bibinfo {author} {\bibfnamefont {A.}~\bibnamefont {Lucas}},\ }\bibfield  {title} {\bibinfo {title} {Ising formulations of many {NP} problems},\ }\href {https://doi.org/10.3389/fphy.2014.00005} {\bibfield  {journal} {\bibinfo  {journal} {Front. Phys.}\ }\textbf {\bibinfo {volume} {2}},\ \bibinfo {pages} {5} (\bibinfo {year} {2014})}\BibitemShut {NoStop}%
\bibitem [{\citenamefont {Blinder}\ \emph {et~al.}(2017)\citenamefont {Blinder}, \citenamefont {Dupont}, \citenamefont {Mukhopadhyay}, \citenamefont {Grbi\ifmmode~\acute{c}\else \'{c}\fi{}}, \citenamefont {Laflorencie}, \citenamefont {Capponi}, \citenamefont {Mayaffre}, \citenamefont {Berthier}, \citenamefont {Paduan-Filho},\ and\ \citenamefont {Horvati\ifmmode~\acute{c}\else \'{c}\fi{}}}]{PhysRevB.95.020404}%
  \BibitemOpen
  \bibfield  {author} {\bibinfo {author} {\bibfnamefont {R.}~\bibnamefont {Blinder}}, \bibinfo {author} {\bibfnamefont {M.}~\bibnamefont {Dupont}}, \bibinfo {author} {\bibfnamefont {S.}~\bibnamefont {Mukhopadhyay}}, \bibinfo {author} {\bibfnamefont {M.~S.}\ \bibnamefont {Grbi\ifmmode~\acute{c}\else \'{c}\fi{}}}, \bibinfo {author} {\bibfnamefont {N.}~\bibnamefont {Laflorencie}}, \bibinfo {author} {\bibfnamefont {S.}~\bibnamefont {Capponi}}, \bibinfo {author} {\bibfnamefont {H.}~\bibnamefont {Mayaffre}}, \bibinfo {author} {\bibfnamefont {C.}~\bibnamefont {Berthier}}, \bibinfo {author} {\bibfnamefont {A.}~\bibnamefont {Paduan-Filho}},\ and\ \bibinfo {author} {\bibfnamefont {M.}~\bibnamefont {Horvati\ifmmode~\acute{c}\else \'{c}\fi{}}},\ }\bibfield  {title} {\bibinfo {title} {Nuclear magnetic resonance study of the magnetic-field-induced ordered phase in the ${\text{nicl}}_{2}\text{\ensuremath{-}}4\text{SC}{({\text{NH}}_{2})}_{2}$ compound},\ }\href {https://doi.org/10.1103/PhysRevB.95.020404} {\bibfield
  {journal} {\bibinfo  {journal} {Phys. Rev. B}\ }\textbf {\bibinfo {volume} {95}},\ \bibinfo {pages} {020404} (\bibinfo {year} {2017})}\BibitemShut {NoStop}%
\bibitem [{\citenamefont {Furuya}\ \emph {et~al.}(2016)\citenamefont {Furuya}, \citenamefont {Dupont}, \citenamefont {Capponi}, \citenamefont {Laflorencie},\ and\ \citenamefont {Giamarchi}}]{PhysRevB.94.144403}%
  \BibitemOpen
  \bibfield  {author} {\bibinfo {author} {\bibfnamefont {S.~C.}\ \bibnamefont {Furuya}}, \bibinfo {author} {\bibfnamefont {M.}~\bibnamefont {Dupont}}, \bibinfo {author} {\bibfnamefont {S.}~\bibnamefont {Capponi}}, \bibinfo {author} {\bibfnamefont {N.}~\bibnamefont {Laflorencie}},\ and\ \bibinfo {author} {\bibfnamefont {T.}~\bibnamefont {Giamarchi}},\ }\bibfield  {title} {\bibinfo {title} {Dimensional modulation of spontaneous magnetic order in quasi-two-dimensional quantum antiferromagnets},\ }\href {https://doi.org/10.1103/PhysRevB.94.144403} {\bibfield  {journal} {\bibinfo  {journal} {Phys. Rev. B}\ }\textbf {\bibinfo {volume} {94}},\ \bibinfo {pages} {144403} (\bibinfo {year} {2016})}\BibitemShut {NoStop}%
\bibitem [{\citenamefont {Bollmark}\ \emph {et~al.}(2020)\citenamefont {Bollmark}, \citenamefont {Laflorencie},\ and\ \citenamefont {Kantian}}]{PhysRevB.102.195145}%
  \BibitemOpen
  \bibfield  {author} {\bibinfo {author} {\bibfnamefont {G.}~\bibnamefont {Bollmark}}, \bibinfo {author} {\bibfnamefont {N.}~\bibnamefont {Laflorencie}},\ and\ \bibinfo {author} {\bibfnamefont {A.}~\bibnamefont {Kantian}},\ }\bibfield  {title} {\bibinfo {title} {Dimensional crossover and phase transitions in coupled chains: Density matrix renormalization group results},\ }\href {https://doi.org/10.1103/PhysRevB.102.195145} {\bibfield  {journal} {\bibinfo  {journal} {Phys. Rev. B}\ }\textbf {\bibinfo {volume} {102}},\ \bibinfo {pages} {195145} (\bibinfo {year} {2020})}\BibitemShut {NoStop}%
\bibitem [{\citenamefont {Bollmark}\ \emph {et~al.}(2023)\citenamefont {Bollmark}, \citenamefont {K\"ohler}, \citenamefont {Pizzino}, \citenamefont {Yang}, \citenamefont {Hofmann}, \citenamefont {Shi}, \citenamefont {Zhang}, \citenamefont {Giamarchi},\ and\ \citenamefont {Kantian}}]{PhysRevX.13.011039}%
  \BibitemOpen
  \bibfield  {author} {\bibinfo {author} {\bibfnamefont {G.}~\bibnamefont {Bollmark}}, \bibinfo {author} {\bibfnamefont {T.}~\bibnamefont {K\"ohler}}, \bibinfo {author} {\bibfnamefont {L.}~\bibnamefont {Pizzino}}, \bibinfo {author} {\bibfnamefont {Y.}~\bibnamefont {Yang}}, \bibinfo {author} {\bibfnamefont {J.~S.}\ \bibnamefont {Hofmann}}, \bibinfo {author} {\bibfnamefont {H.}~\bibnamefont {Shi}}, \bibinfo {author} {\bibfnamefont {S.}~\bibnamefont {Zhang}}, \bibinfo {author} {\bibfnamefont {T.}~\bibnamefont {Giamarchi}},\ and\ \bibinfo {author} {\bibfnamefont {A.}~\bibnamefont {Kantian}},\ }\bibfield  {title} {\bibinfo {title} {Solving 2d and 3d lattice models of correlated fermions---combining matrix product states with mean-field theory},\ }\href {https://doi.org/10.1103/PhysRevX.13.011039} {\bibfield  {journal} {\bibinfo  {journal} {Phys. Rev. X}\ }\textbf {\bibinfo {volume} {13}},\ \bibinfo {pages} {011039} (\bibinfo {year} {2023})}\BibitemShut {NoStop}%
\bibitem [{\citenamefont {Bollmark}\ \emph {et~al.}(2025)\citenamefont {Bollmark}, \citenamefont {K\"ohler},\ and\ \citenamefont {Kantian}}]{PhysRevB.111.125141}%
  \BibitemOpen
  \bibfield  {author} {\bibinfo {author} {\bibfnamefont {G.}~\bibnamefont {Bollmark}}, \bibinfo {author} {\bibfnamefont {T.}~\bibnamefont {K\"ohler}},\ and\ \bibinfo {author} {\bibfnamefont {A.}~\bibnamefont {Kantian}},\ }\bibfield  {title} {\bibinfo {title} {Resolving competition of charge density wave and superconducting phases using the matrix product state plus mean field algorithm},\ }\href {https://doi.org/10.1103/PhysRevB.111.125141} {\bibfield  {journal} {\bibinfo  {journal} {Phys. Rev. B}\ }\textbf {\bibinfo {volume} {111}},\ \bibinfo {pages} {125141} (\bibinfo {year} {2025})}\BibitemShut {NoStop}%
\bibitem [{\citenamefont {Sherrington}\ and\ \citenamefont {Kirkpatrick}(1975)}]{PhysRevLett.35.1792}%
  \BibitemOpen
  \bibfield  {author} {\bibinfo {author} {\bibfnamefont {D.}~\bibnamefont {Sherrington}}\ and\ \bibinfo {author} {\bibfnamefont {S.}~\bibnamefont {Kirkpatrick}},\ }\bibfield  {title} {\bibinfo {title} {Solvable {M}odel of a {S}pin-{G}lass},\ }\href {https://doi.org/10.1103/PhysRevLett.35.1792} {\bibfield  {journal} {\bibinfo  {journal} {Phys. Rev. Lett.}\ }\textbf {\bibinfo {volume} {35}},\ \bibinfo {pages} {1792} (\bibinfo {year} {1975})}\BibitemShut {NoStop}%
\bibitem [{\citenamefont {Kuhl}\ \emph {et~al.}(1984)\citenamefont {Kuhl}, \citenamefont {Crippen},\ and\ \citenamefont {Friesen}}]{Kuhl1984}%
  \BibitemOpen
  \bibfield  {author} {\bibinfo {author} {\bibfnamefont {F.~S.}\ \bibnamefont {Kuhl}}, \bibinfo {author} {\bibfnamefont {G.~M.}\ \bibnamefont {Crippen}},\ and\ \bibinfo {author} {\bibfnamefont {D.~K.}\ \bibnamefont {Friesen}},\ }\bibfield  {title} {\bibinfo {title} {A combinatorial algorithm for calculating ligand binding},\ }\href {https://doi.org/https://doi.org/10.1002/jcc.540050105} {\bibfield  {journal} {\bibinfo  {journal} {J. Comput. Chem.}\ }\textbf {\bibinfo {volume} {5}},\ \bibinfo {pages} {24} (\bibinfo {year} {1984})}\BibitemShut {NoStop}%
\bibitem [{Note1()}]{Note1}%
  \BibitemOpen
  \bibinfo {note} {The condition $\protect \boldsymbol {h}\neq \protect \boldsymbol {0}$ implies that Eq.~\protect \eqref {eq:obj_function} does not possess a global $\protect \mathbb {Z}_2$ spin-flip symmetry; i.e., the transformation $\protect \hat {Z}_i\to -\protect \hat {Z}_i$ for all $i$ does not leave the Hamiltonian invariant. If $\protect \boldsymbol {h}=\protect \boldsymbol {0}$, one can substitute one of the Pauli-$Z$ operators with one of its eigenvalues (e.g., $\protect \hat {Z}_N\to \pm 1$). This recasts the problem over $N-1$ degrees of freedom with effective fields $h_i=\pm \protect \mathsf {W}_{iN}$ for $i=1,2,\protect \ldots , N-1$, leaving the Hamiltonian's spectrum invariant.}\BibitemShut {Stop}%
\bibitem [{\citenamefont {Tan}\ \emph {et~al.}(2021)\citenamefont {Tan}, \citenamefont {Lemonde}, \citenamefont {Thanasilp}, \citenamefont {Tangpanitanon},\ and\ \citenamefont {Angelakis}}]{tan2021qubit}%
  \BibitemOpen
  \bibfield  {author} {\bibinfo {author} {\bibfnamefont {B.}~\bibnamefont {Tan}}, \bibinfo {author} {\bibfnamefont {M.-A.}\ \bibnamefont {Lemonde}}, \bibinfo {author} {\bibfnamefont {S.}~\bibnamefont {Thanasilp}}, \bibinfo {author} {\bibfnamefont {J.}~\bibnamefont {Tangpanitanon}},\ and\ \bibinfo {author} {\bibfnamefont {D.~G.}\ \bibnamefont {Angelakis}},\ }\bibfield  {title} {\bibinfo {title} {Qubit-efficient encoding schemes for binary optimisation problems},\ }\href@noop {} {\bibfield  {journal} {\bibinfo  {journal} {Quantum}\ }\textbf {\bibinfo {volume} {5}},\ \bibinfo {pages} {454} (\bibinfo {year} {2021})}\BibitemShut {NoStop}%
\bibitem [{\citenamefont {Dupont}\ and\ \citenamefont {Sundar}(2024)}]{PhysRevA.109.012429}%
  \BibitemOpen
  \bibfield  {author} {\bibinfo {author} {\bibfnamefont {M.}~\bibnamefont {Dupont}}\ and\ \bibinfo {author} {\bibfnamefont {B.}~\bibnamefont {Sundar}},\ }\bibfield  {title} {\bibinfo {title} {Extending relax-and-round combinatorial optimization solvers with quantum correlations},\ }\href {https://doi.org/10.1103/PhysRevA.109.012429} {\bibfield  {journal} {\bibinfo  {journal} {Phys. Rev. A}\ }\textbf {\bibinfo {volume} {109}},\ \bibinfo {pages} {012429} (\bibinfo {year} {2024})}\BibitemShut {NoStop}%
\bibitem [{\citenamefont {Gao}\ and\ \citenamefont {Han}(2012)}]{Gao2012}%
  \BibitemOpen
  \bibfield  {author} {\bibinfo {author} {\bibfnamefont {F.}~\bibnamefont {Gao}}\ and\ \bibinfo {author} {\bibfnamefont {L.}~\bibnamefont {Han}},\ }\bibfield  {title} {\bibinfo {title} {{Implementing the Nelder-Mead simplex algorithm with adaptive parameters}},\ }\href {https://doi.org/10.1007/s10589-010-9329-3} {\bibfield  {journal} {\bibinfo  {journal} {Comput. Optim. Appl.}\ }\textbf {\bibinfo {volume} {51}},\ \bibinfo {pages} {259} (\bibinfo {year} {2012})}\BibitemShut {NoStop}%
\bibitem [{\citenamefont {Brandao}\ \emph {et~al.}(2018)\citenamefont {Brandao}, \citenamefont {Broughton}, \citenamefont {Farhi}, \citenamefont {Gutmann},\ and\ \citenamefont {Neven}}]{Brandao2018}%
  \BibitemOpen
  \bibfield  {author} {\bibinfo {author} {\bibfnamefont {F.~G.}\ \bibnamefont {Brandao}}, \bibinfo {author} {\bibfnamefont {M.}~\bibnamefont {Broughton}}, \bibinfo {author} {\bibfnamefont {E.}~\bibnamefont {Farhi}}, \bibinfo {author} {\bibfnamefont {S.}~\bibnamefont {Gutmann}},\ and\ \bibinfo {author} {\bibfnamefont {H.}~\bibnamefont {Neven}},\ }\bibfield  {title} {\bibinfo {title} {{For Fixed Control Parameters the Quantum Approximate Optimization Algorithm's Objective Function Value Concentrates for Typical Instances}},\ }\href {https://arxiv.org/abs/1812.04170} {\bibfield  {journal} {\bibinfo  {journal} {arXiv:1812.04170}\ } (\bibinfo {year} {2018})}\BibitemShut {NoStop}%
\bibitem [{\citenamefont {Farhi}\ \emph {et~al.}(2022)\citenamefont {Farhi}, \citenamefont {Goldstone}, \citenamefont {Gutmann},\ and\ \citenamefont {Zhou}}]{Farhi2022}%
  \BibitemOpen
  \bibfield  {author} {\bibinfo {author} {\bibfnamefont {E.}~\bibnamefont {Farhi}}, \bibinfo {author} {\bibfnamefont {J.}~\bibnamefont {Goldstone}}, \bibinfo {author} {\bibfnamefont {S.}~\bibnamefont {Gutmann}},\ and\ \bibinfo {author} {\bibfnamefont {L.}~\bibnamefont {Zhou}},\ }\bibfield  {title} {\bibinfo {title} {The {Q}uantum {A}pproximate {O}ptimization {A}lgorithm and the {S}herrington-{K}irkpatrick {M}odel at {I}nfinite {S}ize},\ }\href {https://doi.org/10.22331/q-2022-07-07-759} {\bibfield  {journal} {\bibinfo  {journal} {{Quantum}}\ }\textbf {\bibinfo {volume} {6}},\ \bibinfo {pages} {759} (\bibinfo {year} {2022})}\BibitemShut {NoStop}%
\bibitem [{\citenamefont {Shaydulin}\ \emph {et~al.}(2019)\citenamefont {Shaydulin}, \citenamefont {Safro},\ and\ \citenamefont {Larson}}]{8916288}%
  \BibitemOpen
  \bibfield  {author} {\bibinfo {author} {\bibfnamefont {R.}~\bibnamefont {Shaydulin}}, \bibinfo {author} {\bibfnamefont {I.}~\bibnamefont {Safro}},\ and\ \bibinfo {author} {\bibfnamefont {J.}~\bibnamefont {Larson}},\ }\bibfield  {title} {\bibinfo {title} {Multistart methods for quantum approximate optimization},\ }in\ \href {https://doi.org/10.1109/HPEC.2019.8916288} {\emph {\bibinfo {booktitle} {2019 IEEE High Performance Extreme Computing Conference (HPEC)}}}\ (\bibinfo {year} {2019})\ pp.\ \bibinfo {pages} {1--8}\BibitemShut {NoStop}%
\bibitem [{\citenamefont {Zhou}\ \emph {et~al.}(2020)\citenamefont {Zhou}, \citenamefont {Wang}, \citenamefont {Choi}, \citenamefont {Pichler},\ and\ \citenamefont {Lukin}}]{PhysRevX.10.021067}%
  \BibitemOpen
  \bibfield  {author} {\bibinfo {author} {\bibfnamefont {L.}~\bibnamefont {Zhou}}, \bibinfo {author} {\bibfnamefont {S.-T.}\ \bibnamefont {Wang}}, \bibinfo {author} {\bibfnamefont {S.}~\bibnamefont {Choi}}, \bibinfo {author} {\bibfnamefont {H.}~\bibnamefont {Pichler}},\ and\ \bibinfo {author} {\bibfnamefont {M.~D.}\ \bibnamefont {Lukin}},\ }\bibfield  {title} {\bibinfo {title} {Quantum approximate optimization algorithm: Performance, mechanism, and implementation on near-term devices},\ }\href {https://doi.org/10.1103/PhysRevX.10.021067} {\bibfield  {journal} {\bibinfo  {journal} {Phys. Rev. X}\ }\textbf {\bibinfo {volume} {10}},\ \bibinfo {pages} {021067} (\bibinfo {year} {2020})}\BibitemShut {NoStop}%
\bibitem [{\citenamefont {Wurtz}\ and\ \citenamefont {Love}(2021)}]{PhysRevA.103.042612}%
  \BibitemOpen
  \bibfield  {author} {\bibinfo {author} {\bibfnamefont {J.}~\bibnamefont {Wurtz}}\ and\ \bibinfo {author} {\bibfnamefont {P.}~\bibnamefont {Love}},\ }\bibfield  {title} {\bibinfo {title} {Maxcut quantum approximate optimization algorithm performance guarantees for p>1},\ }\href {https://doi.org/10.1103/PhysRevA.103.042612} {\bibfield  {journal} {\bibinfo  {journal} {Phys. Rev. A}\ }\textbf {\bibinfo {volume} {103}},\ \bibinfo {pages} {042612} (\bibinfo {year} {2021})}\BibitemShut {NoStop}%
\bibitem [{\citenamefont {Galda}\ \emph {et~al.}(2021)\citenamefont {Galda}, \citenamefont {Liu}, \citenamefont {Lykov}, \citenamefont {Alexeev},\ and\ \citenamefont {Safro}}]{Galda2021}%
  \BibitemOpen
  \bibfield  {author} {\bibinfo {author} {\bibfnamefont {A.}~\bibnamefont {Galda}}, \bibinfo {author} {\bibfnamefont {X.}~\bibnamefont {Liu}}, \bibinfo {author} {\bibfnamefont {D.}~\bibnamefont {Lykov}}, \bibinfo {author} {\bibfnamefont {Y.}~\bibnamefont {Alexeev}},\ and\ \bibinfo {author} {\bibfnamefont {I.}~\bibnamefont {Safro}},\ }\bibfield  {title} {\bibinfo {title} {Transferability of optimal qaoa parameters between random graphs},\ }in\ \href {https://doi.org/10.1109/QCE52317.2021.00034} {\emph {\bibinfo {booktitle} {2021 IEEE International Conference on Quantum Computing and Engineering (QCE)}}}\ (\bibinfo {year} {2021})\ pp.\ \bibinfo {pages} {171--180}\BibitemShut {NoStop}%
\bibitem [{\citenamefont {Galda}\ \emph {et~al.}(2023)\citenamefont {Galda}, \citenamefont {Gupta}, \citenamefont {Falla}, \citenamefont {Liu}, \citenamefont {Lykov}, \citenamefont {Alexeev},\ and\ \citenamefont {Safro}}]{Galda2023}%
  \BibitemOpen
  \bibfield  {author} {\bibinfo {author} {\bibfnamefont {A.}~\bibnamefont {Galda}}, \bibinfo {author} {\bibfnamefont {E.}~\bibnamefont {Gupta}}, \bibinfo {author} {\bibfnamefont {J.}~\bibnamefont {Falla}}, \bibinfo {author} {\bibfnamefont {X.}~\bibnamefont {Liu}}, \bibinfo {author} {\bibfnamefont {D.}~\bibnamefont {Lykov}}, \bibinfo {author} {\bibfnamefont {Y.}~\bibnamefont {Alexeev}},\ and\ \bibinfo {author} {\bibfnamefont {I.}~\bibnamefont {Safro}},\ }\bibfield  {title} {\bibinfo {title} {Similarity-based parameter transferability in the quantum approximate optimization algorithm},\ }\bibfield  {journal} {\bibinfo  {journal} {Front. Quantum Sci. Technol.}\ }\textbf {\bibinfo {volume} {Volume 2 - 2023}},\ \href {https://doi.org/10.3389/frqst.2023.1200975} {10.3389/frqst.2023.1200975} (\bibinfo {year} {2023})\BibitemShut {NoStop}%
\bibitem [{\citenamefont {Shaydulin}\ \emph {et~al.}(2023)\citenamefont {Shaydulin}, \citenamefont {Lotshaw}, \citenamefont {Larson}, \citenamefont {Ostrowski},\ and\ \citenamefont {Humble}}]{10.1145/3584706}%
  \BibitemOpen
  \bibfield  {author} {\bibinfo {author} {\bibfnamefont {R.}~\bibnamefont {Shaydulin}}, \bibinfo {author} {\bibfnamefont {P.~C.}\ \bibnamefont {Lotshaw}}, \bibinfo {author} {\bibfnamefont {J.}~\bibnamefont {Larson}}, \bibinfo {author} {\bibfnamefont {J.}~\bibnamefont {Ostrowski}},\ and\ \bibinfo {author} {\bibfnamefont {T.~S.}\ \bibnamefont {Humble}},\ }\bibfield  {title} {\bibinfo {title} {Parameter transfer for quantum approximate optimization of weighted maxcut},\ }\bibfield  {journal} {\bibinfo  {journal} {ACM Transactions on Quantum Computing}\ }\textbf {\bibinfo {volume} {4}},\ \href {https://doi.org/10.1145/3584706} {10.1145/3584706} (\bibinfo {year} {2023})\BibitemShut {NoStop}%
\bibitem [{\citenamefont {Moré}\ \emph {et~al.}(1980)\citenamefont {Moré}, \citenamefont {Garbow},\ and\ \citenamefont {Hillstrom}}]{MoreGarbowHillstrom1980}%
  \BibitemOpen
  \bibfield  {author} {\bibinfo {author} {\bibfnamefont {J.~J.}\ \bibnamefont {Moré}}, \bibinfo {author} {\bibfnamefont {B.~S.}\ \bibnamefont {Garbow}},\ and\ \bibinfo {author} {\bibfnamefont {K.~E.}\ \bibnamefont {Hillstrom}},\ }\href {https://doi.org/10.2172/6997568} {\emph {\bibinfo {title} {{User Guide for MINPACK-1}}}},\ \bibinfo {type} {Tech. Rep.}\ (\bibinfo  {institution} {Argonne National Laboratory},\ \bibinfo {address} {Argonne, Illinois, USA},\ \bibinfo {year} {1980})\BibitemShut {NoStop}%
\bibitem [{\citenamefont {Parisi}(1983)}]{PhysRevLett.50.1946}%
  \BibitemOpen
  \bibfield  {author} {\bibinfo {author} {\bibfnamefont {G.}~\bibnamefont {Parisi}},\ }\bibfield  {title} {\bibinfo {title} {Order parameter for spin-glasses},\ }\href {https://doi.org/10.1103/PhysRevLett.50.1946} {\bibfield  {journal} {\bibinfo  {journal} {Phys. Rev. Lett.}\ }\textbf {\bibinfo {volume} {50}},\ \bibinfo {pages} {1946} (\bibinfo {year} {1983})}\BibitemShut {NoStop}%
\bibitem [{Note2()}]{Note2}%
  \BibitemOpen
  \bibinfo {note} {Under depth-$p$ QAOA, the average energy for SK problems (Eq.~\protect \eqref {eq:obj_function}) scales asymptotically with system size $N$ as $\langle \protect \hat {C}\rangle _p\sim \alpha _p N^{3/2}$, where $\alpha _p\sim O(1)$ is the energy density~\cite {Farhi2022}. Upon decomposition, the problem splits into $K$ independent subproblems, each effectively an SK instance with $N/K$ variables. Consequently, the combined energy of the factorized system (Eq.~\protect \eqref {eq:factorized_objective_function}) scales as $\langle \protect \hat {\protect \tilde {C}}\rangle \sim K\alpha _p(N/K)^{3/2}$. This yields a relative energy scaling of $\langle \protect \hat {\protect \tilde {C}}\rangle \sim \langle \protect \hat {C}\rangle _p/\protect \sqrt {K}$. For the specific case of $p=1$, the density is given by $\alpha _{1}=-1/\protect \sqrt {4e}$~\cite {Farhi2022}.}\BibitemShut {Stop}%
\bibitem [{\citenamefont {Dupont}\ \emph {et~al.}(2023)\citenamefont {Dupont}, \citenamefont {Evert}, \citenamefont {Hodson}, \citenamefont {Sundar}, \citenamefont {Jeffrey}, \citenamefont {Yamaguchi}, \citenamefont {Feng}, \citenamefont {Maciejewski}, \citenamefont {Hadfield}, \citenamefont {Alam}, \citenamefont {Wang}, \citenamefont {Grabbe}, \citenamefont {Lott}, \citenamefont {Rieffel}, \citenamefont {Venturelli},\ and\ \citenamefont {Reagor}}]{Dupont2023}%
  \BibitemOpen
  \bibfield  {author} {\bibinfo {author} {\bibfnamefont {M.}~\bibnamefont {Dupont}}, \bibinfo {author} {\bibfnamefont {B.}~\bibnamefont {Evert}}, \bibinfo {author} {\bibfnamefont {M.~J.}\ \bibnamefont {Hodson}}, \bibinfo {author} {\bibfnamefont {B.}~\bibnamefont {Sundar}}, \bibinfo {author} {\bibfnamefont {S.}~\bibnamefont {Jeffrey}}, \bibinfo {author} {\bibfnamefont {Y.}~\bibnamefont {Yamaguchi}}, \bibinfo {author} {\bibfnamefont {D.}~\bibnamefont {Feng}}, \bibinfo {author} {\bibfnamefont {F.~B.}\ \bibnamefont {Maciejewski}}, \bibinfo {author} {\bibfnamefont {S.}~\bibnamefont {Hadfield}}, \bibinfo {author} {\bibfnamefont {M.~S.}\ \bibnamefont {Alam}}, \bibinfo {author} {\bibfnamefont {Z.}~\bibnamefont {Wang}}, \bibinfo {author} {\bibfnamefont {S.}~\bibnamefont {Grabbe}}, \bibinfo {author} {\bibfnamefont {P.~A.}\ \bibnamefont {Lott}}, \bibinfo {author} {\bibfnamefont {E.~G.}\ \bibnamefont {Rieffel}}, \bibinfo {author} {\bibfnamefont {D.}~\bibnamefont {Venturelli}},\ and\ \bibinfo {author} {\bibfnamefont
  {M.~J.}\ \bibnamefont {Reagor}},\ }\bibfield  {title} {\bibinfo {title} {Quantum-enhanced greedy combinatorial optimization solver},\ }\href {https://doi.org/10.1126/sciadv.adi0487} {\bibfield  {journal} {\bibinfo  {journal} {Sci. Adv.}\ }\textbf {\bibinfo {volume} {9}},\ \bibinfo {pages} {eadi0487} (\bibinfo {year} {2023})}\BibitemShut {NoStop}%
\bibitem [{\citenamefont {Aizenman}\ \emph {et~al.}(1987)\citenamefont {Aizenman}, \citenamefont {Lebowitz},\ and\ \citenamefont {Ruelle}}]{Aizenman1987}%
  \BibitemOpen
  \bibfield  {author} {\bibinfo {author} {\bibfnamefont {M.}~\bibnamefont {Aizenman}}, \bibinfo {author} {\bibfnamefont {J.~L.}\ \bibnamefont {Lebowitz}},\ and\ \bibinfo {author} {\bibfnamefont {D.}~\bibnamefont {Ruelle}},\ }\bibfield  {title} {\bibinfo {title} {Some rigorous results on the {S}herrington-{K}irkpatrick spin glass model},\ }\href {https://doi.org/10.1007/BF01217677} {\bibfield  {journal} {\bibinfo  {journal} {Commun. Math. Phys.}\ }\textbf {\bibinfo {volume} {112}},\ \bibinfo {pages} {3} (\bibinfo {year} {1987})}\BibitemShut {NoStop}%
\bibitem [{\citenamefont {Wurtz}\ and\ \citenamefont {Love}(2022{\natexlab{a}})}]{Wurtz2022}%
  \BibitemOpen
  \bibfield  {author} {\bibinfo {author} {\bibfnamefont {J.}~\bibnamefont {Wurtz}}\ and\ \bibinfo {author} {\bibfnamefont {P.~J.}\ \bibnamefont {Love}},\ }\bibfield  {title} {\bibinfo {title} {Counterdiabaticity and the quantum approximate optimization algorithm},\ }\href {https://doi.org/10.22331/q-2022-01-27-635} {\bibfield  {journal} {\bibinfo  {journal} {{Quantum}}\ }\textbf {\bibinfo {volume} {6}},\ \bibinfo {pages} {635} (\bibinfo {year} {2022}{\natexlab{a}})}\BibitemShut {NoStop}%
\bibitem [{\citenamefont {Boulebnane}\ \emph {et~al.}(2025)\citenamefont {Boulebnane}, \citenamefont {Khan}, \citenamefont {Liu}, \citenamefont {Larson}, \citenamefont {Herman}, \citenamefont {Shaydulin},\ and\ \citenamefont {Pistoia}}]{Boulebnane2025}%
  \BibitemOpen
  \bibfield  {author} {\bibinfo {author} {\bibfnamefont {S.}~\bibnamefont {Boulebnane}}, \bibinfo {author} {\bibfnamefont {A.}~\bibnamefont {Khan}}, \bibinfo {author} {\bibfnamefont {M.}~\bibnamefont {Liu}}, \bibinfo {author} {\bibfnamefont {J.}~\bibnamefont {Larson}}, \bibinfo {author} {\bibfnamefont {D.}~\bibnamefont {Herman}}, \bibinfo {author} {\bibfnamefont {R.}~\bibnamefont {Shaydulin}},\ and\ \bibinfo {author} {\bibfnamefont {M.}~\bibnamefont {Pistoia}},\ }\bibfield  {title} {\bibinfo {title} {{Evidence that the Quantum Approximate Optimization Algorithm Optimizes the Sherrington-Kirkpatrick Model Efficiently in the Average Case}},\ }\href {https://arxiv.org/abs/2505.07929} {\bibfield  {journal} {\bibinfo  {journal} {arXiv:2505.07929}\ } (\bibinfo {year} {2025})}\BibitemShut {NoStop}%
\bibitem [{\citenamefont {Matsuo}\ \emph {et~al.}(2023)\citenamefont {Matsuo}, \citenamefont {Yamashita},\ and\ \citenamefont {Egger}}]{MATSUO20232022EAP1159}%
  \BibitemOpen
  \bibfield  {author} {\bibinfo {author} {\bibfnamefont {A.}~\bibnamefont {Matsuo}}, \bibinfo {author} {\bibfnamefont {S.}~\bibnamefont {Yamashita}},\ and\ \bibinfo {author} {\bibfnamefont {D.~J.}\ \bibnamefont {Egger}},\ }\bibfield  {title} {\bibinfo {title} {A {SAT} {A}pproach to the {I}nitial {M}apping {P}roblem in {S}wap {G}ate {I}nsertion for {C}ommuting {G}ates},\ }\href {https://doi.org/10.1587/transfun.2022EAP1159} {\bibfield  {journal} {\bibinfo  {journal} {IEICE Trans. Fundam. Electron. Commun. Comput. Sci.}\ }\textbf {\bibinfo {volume} {E106.A}},\ \bibinfo {pages} {1424} (\bibinfo {year} {2023})}\BibitemShut {NoStop}%
\bibitem [{\citenamefont {Kirkpatrick}\ \emph {et~al.}(1983)\citenamefont {Kirkpatrick}, \citenamefont {Gelatt},\ and\ \citenamefont {Vecchi}}]{kirkpatrick1983optimization}%
  \BibitemOpen
  \bibfield  {author} {\bibinfo {author} {\bibfnamefont {S.}~\bibnamefont {Kirkpatrick}}, \bibinfo {author} {\bibfnamefont {C.~D.}\ \bibnamefont {Gelatt}},\ and\ \bibinfo {author} {\bibfnamefont {M.~P.}\ \bibnamefont {Vecchi}},\ }\bibfield  {title} {\bibinfo {title} {Optimization by {S}imulated {A}nnealing},\ }\href {https://doi.org/10.1126/science.220.4598.671} {\bibfield  {journal} {\bibinfo  {journal} {Science}\ }\textbf {\bibinfo {volume} {220}},\ \bibinfo {pages} {671} (\bibinfo {year} {1983})}\BibitemShut {NoStop}%
\bibitem [{\citenamefont {Banchi}\ \emph {et~al.}(2020)\citenamefont {Banchi}, \citenamefont {Fingerhuth}, \citenamefont {Babej}, \citenamefont {Ing},\ and\ \citenamefont {Arrazola}}]{doi:10.1126/sciadv.aax1950}%
  \BibitemOpen
  \bibfield  {author} {\bibinfo {author} {\bibfnamefont {L.}~\bibnamefont {Banchi}}, \bibinfo {author} {\bibfnamefont {M.}~\bibnamefont {Fingerhuth}}, \bibinfo {author} {\bibfnamefont {T.}~\bibnamefont {Babej}}, \bibinfo {author} {\bibfnamefont {C.}~\bibnamefont {Ing}},\ and\ \bibinfo {author} {\bibfnamefont {J.~M.}\ \bibnamefont {Arrazola}},\ }\bibfield  {title} {\bibinfo {title} {Molecular docking with gaussian boson sampling},\ }\href {https://doi.org/10.1126/sciadv.aax1950} {\bibfield  {journal} {\bibinfo  {journal} {Sci. Adv.}\ }\textbf {\bibinfo {volume} {6}},\ \bibinfo {pages} {eaax1950} (\bibinfo {year} {2020})}\BibitemShut {NoStop}%
\bibitem [{\citenamefont {Hamilton}\ \emph {et~al.}(2017)\citenamefont {Hamilton}, \citenamefont {Kruse}, \citenamefont {Sansoni}, \citenamefont {Barkhofen}, \citenamefont {Silberhorn},\ and\ \citenamefont {Jex}}]{hamilton2017gaussian}%
  \BibitemOpen
  \bibfield  {author} {\bibinfo {author} {\bibfnamefont {C.~S.}\ \bibnamefont {Hamilton}}, \bibinfo {author} {\bibfnamefont {R.}~\bibnamefont {Kruse}}, \bibinfo {author} {\bibfnamefont {L.}~\bibnamefont {Sansoni}}, \bibinfo {author} {\bibfnamefont {S.}~\bibnamefont {Barkhofen}}, \bibinfo {author} {\bibfnamefont {C.}~\bibnamefont {Silberhorn}},\ and\ \bibinfo {author} {\bibfnamefont {I.}~\bibnamefont {Jex}},\ }\bibfield  {title} {\bibinfo {title} {Gaussian boson sampling},\ }\href {https://doi.org/10.1103/PhysRevLett.119.170501} {\bibfield  {journal} {\bibinfo  {journal} {Phys. Rev. Lett.}\ }\textbf {\bibinfo {volume} {119}},\ \bibinfo {pages} {170501} (\bibinfo {year} {2017})}\BibitemShut {NoStop}%
\bibitem [{\citenamefont {Ding}\ \emph {et~al.}(2024)\citenamefont {Ding}, \citenamefont {Huang},\ and\ \citenamefont {Yuan}}]{PhysRevApplied.21.034036}%
  \BibitemOpen
  \bibfield  {author} {\bibinfo {author} {\bibfnamefont {Q.-M.}\ \bibnamefont {Ding}}, \bibinfo {author} {\bibfnamefont {Y.-M.}\ \bibnamefont {Huang}},\ and\ \bibinfo {author} {\bibfnamefont {X.}~\bibnamefont {Yuan}},\ }\bibfield  {title} {\bibinfo {title} {Molecular docking via quantum approximate optimization algorithm},\ }\href {https://doi.org/10.1103/PhysRevApplied.21.034036} {\bibfield  {journal} {\bibinfo  {journal} {Phys. Rev. Appl.}\ }\textbf {\bibinfo {volume} {21}},\ \bibinfo {pages} {034036} (\bibinfo {year} {2024})}\BibitemShut {NoStop}%
\bibitem [{\citenamefont {Papalitsas}\ \emph {et~al.}(2025)\citenamefont {Papalitsas}, \citenamefont {Guan}, \citenamefont {Waghe}, \citenamefont {Liakos}, \citenamefont {Balatsos},\ and\ \citenamefont {Pantazopoulos}}]{Papalitsas2025}%
  \BibitemOpen
  \bibfield  {author} {\bibinfo {author} {\bibfnamefont {C.}~\bibnamefont {Papalitsas}}, \bibinfo {author} {\bibfnamefont {Y.}~\bibnamefont {Guan}}, \bibinfo {author} {\bibfnamefont {S.}~\bibnamefont {Waghe}}, \bibinfo {author} {\bibfnamefont {A.}~\bibnamefont {Liakos}}, \bibinfo {author} {\bibfnamefont {I.}~\bibnamefont {Balatsos}},\ and\ \bibinfo {author} {\bibfnamefont {V.}~\bibnamefont {Pantazopoulos}},\ }\bibfield  {title} {\bibinfo {title} {{Q}uantum {A}pproximate {O}ptimization {A}lgorithms for {M}olecular {D}ocking},\ }\href {https://arxiv.org/abs/2503.04239} {\bibfield  {journal} {\bibinfo  {journal} {arXiv:2503.04239}\ } (\bibinfo {year} {2025})}\BibitemShut {NoStop}%
\bibitem [{\citenamefont {Chandarana}\ \emph {et~al.}(2022)\citenamefont {Chandarana}, \citenamefont {Hegade}, \citenamefont {Paul}, \citenamefont {Albarr{\'a}n-Ar{\'a}mburu},\ and\ \citenamefont {Solano}}]{chandarana2022digitized}%
  \BibitemOpen
  \bibfield  {author} {\bibinfo {author} {\bibfnamefont {P.}~\bibnamefont {Chandarana}}, \bibinfo {author} {\bibfnamefont {N.~N.}\ \bibnamefont {Hegade}}, \bibinfo {author} {\bibfnamefont {K.}~\bibnamefont {Paul}}, \bibinfo {author} {\bibfnamefont {F.}~\bibnamefont {Albarr{\'a}n-Ar{\'a}mburu}},\ and\ \bibinfo {author} {\bibfnamefont {E.}~\bibnamefont {Solano}},\ }\bibfield  {title} {\bibinfo {title} {Digitized-counterdiabatic quantum approximate optimization algorithm},\ }\href {https://doi.org/10.1103/PhysRevResearch.4.013141} {\bibfield  {journal} {\bibinfo  {journal} {Phys. Rev. Res.}\ }\textbf {\bibinfo {volume} {4}},\ \bibinfo {pages} {013141} (\bibinfo {year} {2022})}\BibitemShut {NoStop}%
\bibitem [{\citenamefont {Wurtz}\ and\ \citenamefont {Love}(2022{\natexlab{b}})}]{wurtz2022counterdiabaticity}%
  \BibitemOpen
  \bibfield  {author} {\bibinfo {author} {\bibfnamefont {J.}~\bibnamefont {Wurtz}}\ and\ \bibinfo {author} {\bibfnamefont {P.~J.}\ \bibnamefont {Love}},\ }\bibfield  {title} {\bibinfo {title} {Counterdiabaticity and the quantum approximate optimization algorithm},\ }\href {https://doi.org/10.22331/q-2022-02-07-635} {\bibfield  {journal} {\bibinfo  {journal} {Quantum}\ }\textbf {\bibinfo {volume} {6}},\ \bibinfo {pages} {635} (\bibinfo {year} {2022}{\natexlab{b}})}\BibitemShut {NoStop}%
\bibitem [{\citenamefont {Giurgica-Tiron}\ \emph {et~al.}(2020)\citenamefont {Giurgica-Tiron}, \citenamefont {Hindy}, \citenamefont {LaRose}, \citenamefont {Mari},\ and\ \citenamefont {Zeng}}]{9259940}%
  \BibitemOpen
  \bibfield  {author} {\bibinfo {author} {\bibfnamefont {T.}~\bibnamefont {Giurgica-Tiron}}, \bibinfo {author} {\bibfnamefont {Y.}~\bibnamefont {Hindy}}, \bibinfo {author} {\bibfnamefont {R.}~\bibnamefont {LaRose}}, \bibinfo {author} {\bibfnamefont {A.}~\bibnamefont {Mari}},\ and\ \bibinfo {author} {\bibfnamefont {W.~J.}\ \bibnamefont {Zeng}},\ }\bibfield  {title} {\bibinfo {title} {Digital zero noise extrapolation for quantum error mitigation},\ }in\ \href {https://doi.org/10.1109/QCE49297.2020.00045} {\emph {\bibinfo {booktitle} {2020 IEEE International Conference on Quantum Computing and Engineering (QCE)}}}\ (\bibinfo {year} {2020})\ pp.\ \bibinfo {pages} {306--316}\BibitemShut {NoStop}%
\bibitem [{\citenamefont {Bauer}\ \emph {et~al.}(2020)\citenamefont {Bauer}, \citenamefont {Bravyi}, \citenamefont {Motta},\ and\ \citenamefont {Chan}}]{bauer2020quantum}%
  \BibitemOpen
  \bibfield  {author} {\bibinfo {author} {\bibfnamefont {B.}~\bibnamefont {Bauer}}, \bibinfo {author} {\bibfnamefont {S.}~\bibnamefont {Bravyi}}, \bibinfo {author} {\bibfnamefont {M.}~\bibnamefont {Motta}},\ and\ \bibinfo {author} {\bibfnamefont {G.~K.-L.}\ \bibnamefont {Chan}},\ }\bibfield  {title} {\bibinfo {title} {Quantum algorithms for quantum chemistry and quantum materials science},\ }\href {https://doi.org/10.1021/acs.chemrev.9b00829} {\bibfield  {journal} {\bibinfo  {journal} {Chem. Rev.}\ }\textbf {\bibinfo {volume} {120}},\ \bibinfo {pages} {12685} (\bibinfo {year} {2020})}\BibitemShut {NoStop}%
\bibitem [{\citenamefont {Cong}\ \emph {et~al.}(2019)\citenamefont {Cong}, \citenamefont {Choi},\ and\ \citenamefont {Lukin}}]{cong2019quantum}%
  \BibitemOpen
  \bibfield  {author} {\bibinfo {author} {\bibfnamefont {I.}~\bibnamefont {Cong}}, \bibinfo {author} {\bibfnamefont {S.}~\bibnamefont {Choi}},\ and\ \bibinfo {author} {\bibfnamefont {M.~D.}\ \bibnamefont {Lukin}},\ }\bibfield  {title} {\bibinfo {title} {Quantum convolutional neural networks},\ }\href {https://doi.org/10.1038/s41567-019-0648-8} {\bibfield  {journal} {\bibinfo  {journal} {Nat. Phys.}\ }\textbf {\bibinfo {volume} {15}},\ \bibinfo {pages} {1273} (\bibinfo {year} {2019})}\BibitemShut {NoStop}%
\bibitem [{\citenamefont {States)}(2026)}]{data_availability}%
  \BibitemOpen
  \bibfield  {author} {\bibinfo {author} {\bibfnamefont {R.~C.~U.}\ \bibnamefont {States)}},\ }\bibfield  {title} {\bibinfo {title} {Self-consistent mean-field quantum approximate optimization},\ }\href {https://doi.org/10.5281/zenodo.18172244} {10.5281/zenodo.18172244} (\bibinfo {year} {2026})\BibitemShut {NoStop}%
\bibitem [{\citenamefont {{Govinda Rao}}\ \emph {et~al.}(2007)\citenamefont {{Govinda Rao}}, \citenamefont {Bandarage}, \citenamefont {Wang}, \citenamefont {Come}, \citenamefont {Perola}, \citenamefont {Wei}, \citenamefont {Tian},\ and\ \citenamefont {Saunders}}]{GOVINDARAO20072250}%
  \BibitemOpen
  \bibfield  {author} {\bibinfo {author} {\bibfnamefont {B.}~\bibnamefont {{Govinda Rao}}}, \bibinfo {author} {\bibfnamefont {U.~K.}\ \bibnamefont {Bandarage}}, \bibinfo {author} {\bibfnamefont {T.}~\bibnamefont {Wang}}, \bibinfo {author} {\bibfnamefont {J.~H.}\ \bibnamefont {Come}}, \bibinfo {author} {\bibfnamefont {E.}~\bibnamefont {Perola}}, \bibinfo {author} {\bibfnamefont {Y.}~\bibnamefont {Wei}}, \bibinfo {author} {\bibfnamefont {S.-K.}\ \bibnamefont {Tian}},\ and\ \bibinfo {author} {\bibfnamefont {J.~O.}\ \bibnamefont {Saunders}},\ }\bibfield  {title} {\bibinfo {title} {{Novel thiol-based TACE inhibitors: Rational design, synthesis, and SAR of thiol-containing aryl sulfonamides}},\ }\href {https://doi.org/https://doi.org/10.1016/j.bmcl.2007.01.064} {\bibfield  {journal} {\bibinfo  {journal} {Bioorg. Med. Chem. Lett.}\ }\textbf {\bibinfo {volume} {17}},\ \bibinfo {pages} {2250} (\bibinfo {year} {2007})}\BibitemShut {NoStop}%
\bibitem [{\citenamefont {Zhang}\ \emph {et~al.}(2021)\citenamefont {Zhang}, \citenamefont {Krieger}, \citenamefont {Zhang}, \citenamefont {Kaya}, \citenamefont {Kaynak}, \citenamefont {Mikulska-Ruminska}, \citenamefont {Doruker}, \citenamefont {Li},\ and\ \citenamefont {Bahar}}]{10.1093/bioinformatics/btab187}%
  \BibitemOpen
  \bibfield  {author} {\bibinfo {author} {\bibfnamefont {S.}~\bibnamefont {Zhang}}, \bibinfo {author} {\bibfnamefont {J.~M.}\ \bibnamefont {Krieger}}, \bibinfo {author} {\bibfnamefont {Y.}~\bibnamefont {Zhang}}, \bibinfo {author} {\bibfnamefont {C.}~\bibnamefont {Kaya}}, \bibinfo {author} {\bibfnamefont {B.}~\bibnamefont {Kaynak}}, \bibinfo {author} {\bibfnamefont {K.}~\bibnamefont {Mikulska-Ruminska}}, \bibinfo {author} {\bibfnamefont {P.}~\bibnamefont {Doruker}}, \bibinfo {author} {\bibfnamefont {H.}~\bibnamefont {Li}},\ and\ \bibinfo {author} {\bibfnamefont {I.}~\bibnamefont {Bahar}},\ }\bibfield  {title} {\bibinfo {title} {Prody 2.0: increased scale and scope after 10 years of protein dynamics modelling with python},\ }\href {https://doi.org/10.1093/bioinformatics/btab187} {\bibfield  {journal} {\bibinfo  {journal} {Bioinformatics}\ }\textbf {\bibinfo {volume} {37}},\ \bibinfo {pages} {3657} (\bibinfo {year} {2021})}\BibitemShut {NoStop}%
\bibitem [{RDKit, online()}]{rdkit}%
  \BibitemOpen
  RDKit, online,\ \href@noop {} {\bibinfo {title} {{RDK}it: Open-source cheminformatics}},\ \bibinfo {howpublished} {\url{http://www.rdkit.org}}\BibitemShut {NoStop}%
\end{thebibliography}%

\end{document}